\documentstyle[twocolumn,aps,epsf,floats]{revtex}

\begin{document}

\twocolumn[\hsize\textwidth\columnwidth\hsize\csname %
@twocolumnfalse\endcsname

\title
{Resonant Raman Scattering in Antiferromagnets}
\author{Dirk K. Morr  and Andrey V. Chubukov,$^*$}
\address
{Department of Physics, University of Wisconsin-Madison,
1150 University Ave., Madison, WI 53706\\
$^*$ P.L. Kapitza Institute for Physical Problems, Moscow, Russia}
\date{\today}
\maketitle
\begin{abstract}

Two-magnon Raman scattering provides important information about  
electronic correlations in the insulating parent compounds of  
high-$T_c$ materials. Recent experiments have shown a strong  
dependence of the Raman signal in $B_{1g}$ geometry on the frequency  
of the incoming photon. We present an analytical 
and numerical study of the Raman 
intensity in the resonant regime. It has been previously argued 
by one of us (A.Ch) and D. Frenkel that the most relevant contribution to the
 Raman vertex at resonance is given by the 
triple resonance diagram.
 We derive an expression for the 
Raman intensity in which we simultaneously include the enhancement due to  
the triple resonance and a final state interaction. 
We compute the two-magnon peak height (TMPH) as a 
function of incident frequency and 
find two maxima at $\omega^{(1)}_{res} \approx 2\Delta + 3J$ and 
$\omega^{(2)}_{res} 
\approx 2\Delta + 8J$. We argue that the high-frequency maximum is cut only
by a quasiparticle damping, while the low-frequency maximum has a finite
amplitude even in the absence of damping. We also 
obtain an evolution of the Raman profile 
from an asymmetric form around $\omega^{(1)}_{res}$ to a symmetric form 
around $\omega^{(2)}_{res}$. 
We further show that the TMPH depends on the  
fermionic quasiparticle damping, the next-nearest neighbor hopping term $t^{\prime}$ and the 
corrections to the interaction vertex between light 
and the fermionic current. We discuss our 
results in the context of recent experiments by  Blumberg {\it et al.} on $Sr_2CuO_2Cl_2$ and 
$YBa_2Cu_3O_{6.1}$ and R\"{u}bhausen {\it et al.} on $PrBa_2Cu_3O_7$ and show that the triple 
resonance theory yields a qualitative and to some 
extent also quantitative understanding of the experimental data.
\end{abstract}
\pacs{PACS:74.25.Ha, 74.25.Jb} 

]

\narrowtext

\section{Introduction}

In recent years a lot of efforts have been undertaken
to understand the pairing mechanism
in high-$T_c$ superconductors \cite{And,AndSch,AndPin}.
Some of the existing 
theories consider an effective electron-electron interaction mediated by spin
fluctuations as the source of the pairing mechanism \cite{Pines,Scal}. In the
parent compounds of the high-$T_c$ materials the strong magnetic correlations
lead to the occurrence of antiferromagnetism. 
Two-magnon Raman scattering is a
valuable tool in probing antiferromagnetism and can thus provide important
insight into the nature of the pairing correlations 
\cite{Mer,CotLoc,Sin90,Bren}.

The two-magnon Raman scattering cross section (Raman intensity)
 is proportional to the Golden Rule transition rate 
\cite{HayLou}
\begin{equation}
R= {8 \pi^3 e^4 \over \hbar^3 V^2 \omega_i \omega_f} 
\sum{ |M_R|^2 \delta(\hbar \omega_i-\hbar 
\omega_f +\epsilon_i-\epsilon_f) }
\label{golden}
\end{equation}
where $i$ and $f$ are the initial and final states of the system, 
$\epsilon_{i,f}$ are the 
corresponding energies and 
($\epsilon_i - \epsilon_f$) is the total energy of the two
magnons in the final state. $M_R=<\hat{e}_f^*|M_R|\hat{e}_i>$ 
is the Raman  matrix 
element (Raman vertex), $\hat{e}_i$ and $\hat{e}_f$ 
are the polarization unit vectors
of the incident and outgoing  photons, 
and  the summation runs over all possible 
initial and final electronic states.  

\begin{figure} [t]
\begin{center}
\leavevmode
\epsffile{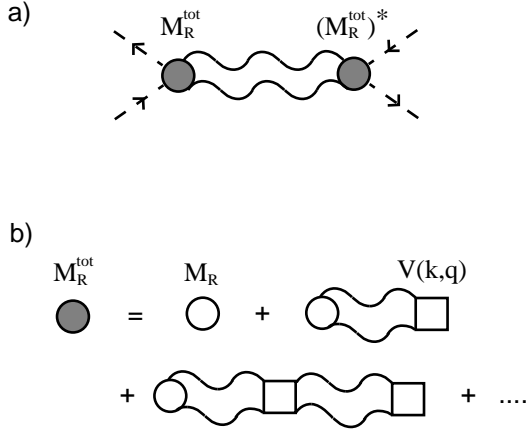}
\end{center}
\caption{{\it (a)} The Raman cross section is 
proportional to this diagram where the intermediate magnons (wavy lines) are on the mass shell. 
The dashed lines present incoming and 
outgoing photons and the filled circles are the full Raman vertices $M_R^{tot}$.
{\it (b)} The full Raman vertices include all effects of multiple magnon-magnon scattering. The 
open circles are the bare Raman vertices $M_R$ and the open squares describe the magnon-magnon 
scattering vertex $V(k,q)$.} 
\label{diaint}
\end{figure}

Graphically, the Raman intensity
is given by the  diagram shown in Fig. \ref{diaint}a, where the intermediate
 magnons (wavy lines) are on the mass shell.
The dashed lines in 
this diagram describe the incident and outgoing photons
and the shaded circles represent the full Raman vertices,
which include all effects of the final state magnon-magnon
interaction \cite{CanGir}.
In conventional Raman experiments, 
one measures the Raman intensity, $R$,
as a function
of transferred photon frequency $\Delta\omega = 
\omega_i - \omega_f$ where $\omega_i$ and $\omega_f$ are the incident and outgoing
photon frequencies, respectively. 
The fingerprint of antiferromagnetism 
in these experiments  is the presence of a two-magnon peak in $R(\Delta \omega)$ \cite{LyoFle}.
In the insulating parent compounds of the high-$T_c$ materials, this peak occurs at a 
transferred frequency of about $3000 cm^{-1}$. 
The two-magnon peak has not only been observed in the insulating compounds, but also 
in electron \cite{TomYos} and hole doped materials \cite{BluLiu}.
 
\begin{figure} [t]
\begin{center}
\leavevmode
\epsffile{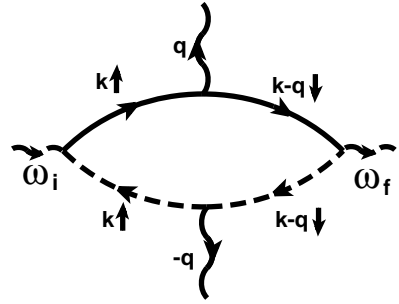}
\end{center}
\caption{The triple resonance diagram which 
yields the dominant contribution to the Raman intensity 
in the resonant regime. Solid and dashed lines represent fermions from
conduction and valence bands, respectively. Notice that this diagram contains
intraband scattering at the fermion-magnon vertices.}
\label{res_diag}
\end{figure}

Theoretically, most of the analytical and
 numerical studies of two-magnon Raman scattering were
performed within the conventional
phenomenological Loudon-Fleury (LF) theory \cite{FleLou}
 which assumes that 
 the matrix element $M_R$ for the interaction between 
photons and magnons is frequency independent. This implies that the theory
neglects the internal structure which the matrix element possesses
since the spin-photon interaction is  actually    
mediated by fermions: the incident photon creates a particle-hole pair which
emits two spin excitations and annihilates into an outgoing photon 
(see e.g., Fig. \ref{res_diag}).
Despite this weakness, the LF theory was
originally considered as a suitable theory for Raman scattering in the 
parent high-$T_c$ materials because it
predicts that  the two-magnon profile should have a peak
at a transferred frequency of about $2.8J$ where $J$ is the magnetic
exchange interaction. A comparison with the data \cite{LyoFle,LyoFle2,SulFle} then 
yields $J = 0.12 eV$ which  is fully consistent
with the value for the in-plane exchange interaction extracted from 
neutron scattering \cite{Hayden} and NMR data \cite{Imai}. 

Recent experiments on single-layer $Sr_2CuO_2Cl_2$ and double-layer
$YBa_2Cu_3O_{6.1}$\cite{BluAbb} as well as on $PrBa_2Cu_3O_7$\cite{RubDie}, 
however,                                                                                   
presented some qualitative features of the Raman signal 
which cannot be explained within the framework of the LF theory. In these
experiments, the Raman intensity was measured both as a function of transferred
frequency at a given incident frequency $\omega_i$ (the two-magnon profile),
and as a function of $\omega_i$ at a fixed
transferred frequency $\Delta \omega 
\approx 2.8J$ at which the 
two-magnon profile exhibits a maximum. In the latter case one in fact measures the
 variation of the two-magnon peak height (TMPH) with $\omega_i$.
The experimental features which are in disagreement with the LF 
theory include:\\
$1)$ A strong dependence of the TMPH 
on $\omega_i$ with two distinct 
maxima  at $\omega_i = \omega^{(1)}_{res} 
\approx 2\Delta + 3J$  and at $\omega_i =  \omega^{(2)}_{res} \approx
2\Delta + 8J$, where $2\Delta \sim 1.7 eV$ is the charge transfer gap \cite{LiuKle}. 
Despite quantitative differences between 
 various compounds, the second maximum in 
all compounds is always stronger than the first one. The LF
theory, on the contrary, predicts that the intensity should only undergo a weak
(logarithmical) enhancement at $\omega_i = 2\Delta$ and $\omega_i = 2\Delta + 2.8J$ (ingoing
and outgoing resonances). No enhancement, however,
has been experimentally observed at $\omega_i = 2\Delta$.\\
$2)$ The shape of the two-magnon profile is asymmetric and possesses a
shoulder-like feature for transferred frequencies above the two-magnon peak, i.e., for 
$\Delta \omega > 2.8J$. This feature 
has been observed around the first resonance at 
$\omega_{res}^{(1)}$; it practically disappears when the frequency 
of the incident photon approaches the second resonance at $\omega_{res}^{(2)}$.

\begin{figure} [t]
\begin{center}
\leavevmode
\epsffile{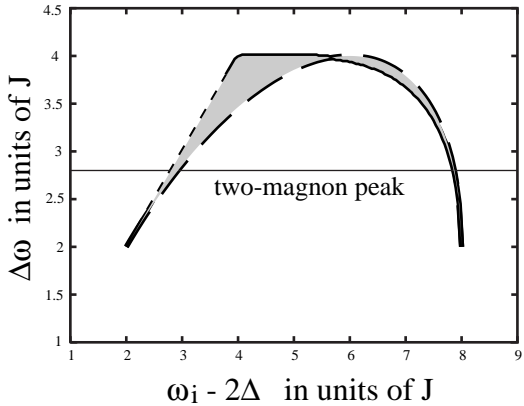}
\end{center}
\caption{The shaded area represent the region in the $(\omega_i, 
\Delta \omega)$ plane 
in which the triple resonance occurs. The horizontal line corresponds to
$\Delta \omega =2.8J$ at which the two-magnon profile has a peak.}
\label{trip_res}
\end{figure}

Motivated by these findings, several groups studied two-magnon Raman scattering
beyond the LF approximation \cite{ShaShr,ChuFre,SchKam}. It has been shown  
that the validity of 
the LF theory is restricted to the nonresonant regime, when the frequency of 
the incident light is much smaller 
than the charge-transfer gap $2 \Delta$ \cite{ShaShr,ChuFre}. Most 
of the experiments, however, are 
performed with photon frequencies slightly above the charge transfer gap. In
this {\it resonant} regime
the internal structure of the Raman matrix element  cannot be neglected. 
Chubukov and Frenkel (hereafter referred to as CF) 
developed a diagrammatic approach to
Raman scattering in the framework of the large-$U$ spin-density-wave (SDW)
approach to the Hubbard model at 
half-filling~\cite{ChuFre}. They identified those diagrams which reproduce the LF 
vertex, and in addition identified a new diagram which is not 
included in the LF theory but yields the 
dominant contribution to the scattering process in the resonant regime.  
This new diagram has the largest  amplitude when $|\omega_{i,f} - 2\Delta| =
O(J)$, and in addition, diverges in the absence of a fermionic damping
in a  small region 
(nearly a single critical line) in the $(\omega_i,
\Delta \omega)$ plane where all three terms in the denominator 
vanish simultaneously (see Fig.~\ref{trip_res}). 
Due to this property, the new diagram identified by CF 
is called the triple resonance diagram.
The inclusion of a fermionic damping eliminates the divergence, but the
Raman matrix element remains strongly peaked along the critical line.
Since the computation of the full Raman intensity with the triple diagram for
the Raman vertex is rather involved, CF
used a semi-phenomenological approach to analyze the dependence of the TMPH on 
$\omega_i$. They 
considered the final state 
magnon-magnon interaction and the triple resonance enhancement separately,
and conjectured that 
the  experimentally observed two maxima in the TMPH 
occur at $\omega_i$ for which the Raman vertex 
resonates at the same transferred frequency $\Delta \omega =2.8J$ 
at which the two-magnon profile has a peak. 
By analyzing  where the resonant line for the Raman vertex 
 crosses $\Delta \omega =
2.8J$, they obtained  two resonance frequencies, 
$\omega^{(1)}_{res} \sim 2.9J + 2\Delta$ and
$\omega^{(2)}_{res} \sim 7.9J + 2\Delta$ which both agree with the 
experimental data.

The analysis in Ref.~\cite{ChuFre}, however, left several issues open. 
First, the validity of the 
semi-phenomenological approach needs to be verified. 
Second, the quantitative behavior of the TMPH as a function of $\omega_i$  
and, in particular, the form of the peaks 
at $\omega^{(1)}_{res}$ and $\omega^{(2)}_{res}$ and their
relative amplitudes
 have not been studied yet.
CF merely conjectured, without performing explicit
calculations, that the resonance  
at $\omega_{res}^{(1)}$ should be weaker than the one at $\omega^{(2)}_{res}$
because near  $\omega_{res}^{(1)}$, there exists a strong 
restriction on the possible directions of the 
magnon momenta which satisfy the resonance condition at a
given magnon energy. No such
restriction exists near  $\omega_{res}^{(2)}$. 
This conjecture also has to be  verified by explicit calculations.
Third, the anisotropy of the two-magnon profile and its evolution 
with varying incident frequency has not been studied. Forth, the
calculations in CF 
were performed in the framework of a mean-field, large $U$,
spin-density-wave (SWD) approach to the Hubbard model with only nearest-neighbor
hopping. This theory, however, possesses the weakness that it predicts 
that the maximum of the valence band is degenerate along the
boundary of the magnetic Brillouin zone. Meanwhile, experiments on
$Sr_2CuO_2Cl_2$ have demonstrated that the valence fermions possess a strong 
dispersion along the magnetic Brillouin zone boundary with maxima 
at $(\pi/2,\pi/2)$  and symmetry related points \cite{Wells,LaR}. This
dispersion can easily be reproduced in the SDW formalism
 if one includes a next-nearest-neighbor hopping, $t^{\prime}$.  This, 
however, changes the
 energy denominator  in the triple resonance diagram, and one
 therefore has to reexamine the conclusions of CF 
by performing their calculations for the
$(t-t^{\prime}-U)$ model.  The inclusion of $t^{\prime}$  is particularly
relevant for computations  near $\omega^{(1)}_{res}$ 
since the dominant
contribution to the Raman vertex in this frequency range comes from
fermions near the top of the valence band whose degeneracy is lifted by
$t^{\prime}$. 

The goal of the present paper is to address the above issues. 
We will compute below the Raman intensity 
including both a final state interaction 
and the enhancement of the Raman vertex due to  the
triple resonance.  We will study the Raman 
profile and the TMPH numerically and analytically and 
demonstrate that the two peaks in the TMPH survive
the effects of the magnon-magnon interaction.
We will analyze the relative amplitude of the TMPH near
 $\omega^{(1)}_{res}$ and 
$\omega^{(2)}_{res}$ and show that although the  
divergent piece  near 
$\omega^{(1)}_{res}$ is much weaker than the one near $\omega^{(2)}_{res}$, 
the nondivergent term is much larger near 
$\omega^{(1)}_{res}$. As a result,  the relative amplitude of the two peaks
in the TMPH turns out to be strongly dependent
 on the quasiparticle damping which cuts the 
divergent part but  does not affect the subleading term substantially. 
We will also show that both peaks in the TMPH are anisotropic - the intensity
drops much faster on the high-frequency side of each of the peaks.
Further, we will 
study the effects on the TMPH of a next-nearest neighbor hopping 
$t^{\prime}$  and vertex corrections to the interaction between light and 
fermionic quasiparticles. 

 We will also study how the two-magnon lineshape evolves with the incident
frequency, and show that it changes from an asymmetric form for 
$\omega_i \geq \omega^{(1)}_{res}$ to a symmetric 
form around $\omega_i \approx \omega^{(2)}_{res}$. 

We will compare our results with the experimental data on
 on $Sr_2CuO_2Cl_2$ 
and $YBa_2Cu_3O_{6.1}$ by Blumberg {\it et al.} and on $PrBa_2Cu_3O_7$ 
by R\"{u}bhausen {\it et al.}  and demonstrate that all
the features in the two-magnon profile and the TMPH observed 
in Raman experiments  can be qualitatively described by the triple 
resonance diagram. At the same time, we will see that
 quantitative agreement with the data is not always  perfect. 

The paper is organized as follows. 
In Sec.~\ref{form}, we present the formalism and the expressions for the
Raman vertex both in the LF approximation and near the resonance. 
In Sec.~\ref{anal} we present
our analytical results for the vertex near $\omega^{(1,2)}_{res}$ and 
demonstrate
that the divergence is much stronger 
near the upper resonance frequency. We also discuss in this section
how the inclusion of a next-nearest-neighbor hopping 
$t^{\prime}$ affects the resonance 
behavior of the Raman vertex. In Sec.~\ref{numerics} we present our
 numerical results for 
{\it (a)} the Raman line shape for different 
incident frequencies $\omega_i$ (Sec.~\ref{shape}) and {\it (b)} for the 
TMPH as a function of $\omega_i$ (Sec.~\ref{tmph}). We then discuss its 
dependence on the fermionic damping and the inclusion of 
$t^{\prime}$ (Sec.~\ref{tprime}).
In Sec.~\ref{vertex} we  consider vertex corrections to the interaction 
between light and the fermionic current.
Finally, in Sec.\ref{discussion} we compare our 
results with the experimental data.

\section{The Formalism}
\label{form}
The two-magnon Raman scattering cross section is given by Eq.(\ref{golden}).
In this paper, we will focus on the $x^\prime y^\prime$ scattering geometry where
most experiments have been performed.  In this geometry,
the polarization unit vectors of the incident and outgoing 
light are both real (linearly polarized light), perpendicular to each other
and directed at $45^o$ to the crystallographic directions, i.e., 
$\hat{e}_i=x^\prime=(\hat{x}+\hat{y})/\sqrt{2}, 
\hat{e}_f=y^\prime=(\hat{x}-\hat{y})/\sqrt{2}$.
Other scattering geometries for linearly polarized light 
are $x^\prime x^\prime$ where $\hat{e}_i=\hat{e}_f=
(\hat{x}+\hat{y})/\sqrt{2}$, $xy$ where
$\hat{e}_i=\hat{x}, \hat{e}_f= \hat{y}$ and $xx$ where
$\hat{e}_i=\hat{x}, \hat{e}_f= \hat{x}$. For circularly polarized light, the
scattering geometries are LL where 
$\hat{e}_i=\hat{e}_f=(\hat{x}+i\hat{y})/\sqrt{2}$, and LR where
$\hat{e}_i=(\hat{x}+i\hat{y})/\sqrt{2}, \hat{e}_f=(\hat{x}-i\hat{y})/\sqrt{2}$.

The $x^\prime y^\prime$ geometry is often referred to as the $B_{1g}$ geometry. Strictly 
speaking the matrix element in the $x^\prime y^\prime$ scattering channel includes 
$B_{1g}$ and $A_{2g}$ components with $M^{B_{1g}}_R=<x|M_R|x>-<y|M_R|y>$ and 
$M^{A_{2g}}_R=<x|M_R|y>-<y|M_R|x>$. In practice, however, the $A_{2g}$ component is always 
negligible, and we will therefore not distinguish between $x^\prime y^\prime$ and pure $B_{1g}$ 
scattering.  
 
As we already discussed in the introduction, we will
consider Raman scattering in the framework of the large-$U$
SDW formalism for the one-band Hubbard model. In this approach, one 
introduces a long-range antiferromagnetic order and 
decouples the electronic dispersion into two subbands of valence and conduction
fermions.
 
The diagrammatic approach to the Raman scattering in the SDW formalism was
developed by CF. An example for the diagrams which contribute to the bare Raman vertex
is shown in Fig.~\ref{res_diag}. 
This diagram contains two types of vertices: one
for the interaction between fermions and light, and one for the
interaction between fermions and magnons.
The interaction with light appears in the SDW theory as a result of the
modulation of the hopping matrix element by the vector potential of the
electromagnetic field. The spin-fermion
vertices can be straightforwardly obtained from the full expression
of the spin susceptibility which in the SDW theory is given by the RPA series
of bubble diagrams. In each of theses bubbles one fermion is from the valence
band and the other is from conduction band.
 
To obtain the Raman intensity, we need to know the full Raman vertex which 
includes a whole series of magnon-magnon interaction events.
It has already been emphasized several times in the literature that 
the dominant contribution to the magnon-magnon scattering comes 
from the region near the magnetic Brillouin zone boundary 
where the antiferromagnetic magnons behave almost as free particles \cite{CanGir,Davis}. 
In this situation,
the only relevant interaction term has two creation and two annihilation magnon
operators:
\begin{equation}
{\cal H}_{int} = -\frac{4J}{N^2} {\sum_k} {\sum_q} \nu_{k-q}
a^{\dagger}_{k} \beta^{\dagger}_{-k}\beta_{-q}a_q.
\end{equation}
We can now decompose the interaction vertex into 
\begin{equation}
\nu_{k-q} = \nu_k \nu_l + {\tilde \nu}_k {\tilde \nu}_l +
{\bar\nu}_k {\bar \nu}_l + {\bar{\tilde \nu}}_k {\bar{\tilde \nu}}_l \; .
\label{scat}
\end{equation} 
where the different symmetry factors are given by
\begin{eqnarray}
\nu_k &=& \frac{1}{2} (\cos k_x + \cos k_y);~~~~ 
\tilde{\nu}_k = \frac{1}{2} (\cos k_x - \cos k_y); \nonumber \\
\bar{\nu}_k &=& \frac{1}{2} (\sin k_x + \sin k_y);~~~~
\bar{\tilde{\nu}}_k = \frac{1}{2} (\sin k_x - \sin k_y). 
\label{eigenf}
\end{eqnarray} 
Before we discuss our calculations for the full Raman intensity at resonance,
we briefly  review the 
calculation of the  Raman intensity in the nonresonant regime when the 
LF theory is valid. In the LF theory, the bare Raman vertex 
(open circle in Fig.~\ref{diaint}b) 
is assumed to be 
independent of the photon
frequencies  while its dependence on the magnon momentum $q$ has 
the form~\cite{ShaShr,ChuFre}
\begin{eqnarray*}
M_R &=& A~\Big[\nu_q 
(e_{ix}e^{*}_{fx} + e_{iy} e^*_{fy}) \\
&&\quad -(e_{ix}e^{*}_{fx} \cos q_x + e_{iy} e^*_{fy} \cos q_y) \Big] \ ,
\end{eqnarray*}
where $A$ is a constant. In the diagrammatic approach, the LF vertex is
obtained by collecting the diagrams with interband scattering at the
magnon-fermion vertices. At  photon frequencies small compared to the SDW gap,
these diagrams have the largest overall factor.
One can easily check that $M_R$ is finite only in the $B_{1g}$ scattering 
geometry and for LR polarized light. 
In both cases we obtain
$M_R = -A \tilde{\nu}_q$. For this particular form of $M_R$,
only the second term 
in Eq.(\ref{scat}) contributes to the magnon-magnon scattering process.
With this simplification, the summation of the ladder series for the full Raman 
intensity can be reduced to solving
an algebraic equation. Doing this
 we obtain for the full Raman intensity in the 
$B_{1g}$ channel 
\begin{equation}
R (\omega) \propto Im \left[ \frac{I}{1 + I/4S}\right] \ ,
\label{R_LF}
\end{equation}
where $S$ is the value of the spin, and 
\begin{equation}
I = \frac{4JS}{N} {\sum_q} \frac{(\cos q_x - \cos q_y)^2}{\Delta \omega -
2 \omega_q + i\delta} 
\label{eff_ver}
\end{equation}
with $\Delta \omega = \omega_i - \omega_f$ and 
$\omega_q = 4JS \sqrt{1 - \nu^2_q}$ is  the magnon dispersion. 
Eqs.(\ref{R_LF}) and (\ref{eff_ver})
yield a two-magnon peak at $\Delta \omega =
2.8J$ (for $S=1/2$), but $R(\omega)$ clearly contains no dependence on the 
incident photon frequency $\omega_i$ \cite{CanGir,FleLou}.\\ 
As mentioned earlier, the LF theory is only valid for small $\omega_i$.
 When $\omega_i$ is comparable to the gap between the conduction and
valence bands (which in the cuprates is the charge transfer gap), it turns out
that diagrams
with intraband scattering at the fermion-magnon vertices (in contrast to 
interband scattering in the LF diagrams) become dominant. 
The most relevant of these diagrams is shown in Fig.~\ref{res_diag}. 
This diagram is 
called the triple resonance diagram because it contains three 
terms in the denominator 
which can all vanish simultaneously if we adjust 
the incident and final photon frequencies.
The analytical expression for this diagram is given by 
\begin{eqnarray}
& & M_R= -{ 4i \over N} {\sum_k}^{\prime}
  {\Big( \displaystyle{\partial \epsilon_k
\over \partial k} e_i \Big) \Big( \displaystyle{\partial \epsilon_{k-q} \over
\partial k} e_f \Big) \Big[ \mu_q \epsilon_{k-q} - \lambda_q \epsilon_k \Big]^2
\over \Big(\omega_i-2E_k+i\Gamma \Big) \Big(\omega_f-2E_{k-q}+i\Gamma \Big) } \nonumber  \\
& &  \quad \times \Bigg\{ {1 \over \Big(\omega_i- \omega 
-E_k-E_{k-q} +i\Gamma \Big)} \nonumber \\
& & \qquad +   
{1 \over \Big(\omega_f+\omega -E_k-E_{k-q}+i\Gamma \Big)}  \Bigg\} \ ,
\label{tripa}
\end{eqnarray}                                              
where  $\omega$ is the external magnon frequency,
\begin{eqnarray*} 
\epsilon_k&=&-2t(\cos{k_x}+\cos{k_y})=-4t\nu_k, \ 
E_k=\sqrt{\Delta^2+\epsilon_k^2} \ , \nonumber\\
\mu_q&=&\Bigg[ {1\over 2} \Bigg( {1 \over \sqrt{1-\nu_q^2}}+1 \Bigg) 
\Bigg]^{1 \over 
2}, \ \lambda_q=\Bigg[ {1\over 2} \Bigg( {1 \over \sqrt{1-\nu_q^2}}-1 \Bigg) 
\Bigg]^{1 \over 2} ,
\end{eqnarray*}
and the prime indicates summation over the magnetic Brillouin zone. 
The $i \Gamma$ term 
represents the fermionic quasi-particle damping which we assume for
simplicity to be independent of momentum. The actual damping, indeed, should
have some momentum dependence, particularly near the top of the valence 
band~\cite{ChuMus,Sin91,AltBre,Lau}.
Out of the three terms in the numerator, 
the first two are the vertex functions for the interaction
between light and fermions, while the third term is the 
product of the two vertices for the
interaction between fermions and magnons.
\begin{figure} [t]
\begin{center}
\leavevmode
\epsffile{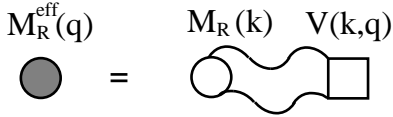}
\end{center}
\caption{The effective 
Raman vertex $M_R^{eff}$ includes a single magnon-magnon scattering event.}
\label{m_eff}
\end{figure}
As follows from Eq.(\ref{tripa}), the resonant Raman vertex $M_R$ 
depends on the magnon momentum, but also, via the denominator, 
on the incident and outgoing photon frequencies and on the magnon frequency.
The frequency dependence of $M_R$ can only be eliminated
in the artificial limit when the fermionic damping is very large and 
overshadows all other
 terms in the denominator.  For small and moderate damping, 
the frequency dependent and also momentum dependent terms 
in the denominator cannot
be ignored, and this makes the computation of the full
Raman intensity rather involved. We found, however, that the 
 diagram which is most difficult to compute is the one without 
a final state interaction. 
At the same time, 
the series of diagrams with at least two scattering events can  
easily be summed up  because the Raman vertex renormalized by the inclusion of
just a single magnon-magnon scattering event no longer depends on the magnon 
frequency while
its dependence on the external magnon momentum $l$  reduces
to a simple $\tilde{\nu}_l$ form for $B_{1g}$ scattering. 
This effective Raman vertex, 
$M^{eff}_R=\tilde{\nu}_l \, \bar{M}^{eff}_R$ 
is shown in Fig.~\ref{m_eff}.

We remind that the experimentally measured Raman profile for any
incident photon frequency contains a prominent two-magnon peak which
 is solely due to magnon-magnon scattering. In this situation, the diagrams
without and with a single magnon-magnon scattering are 
most likely to be less relevant than the diagrams
with multiple scattering events. 
For our analytical considerations, we 
 neglect the diagrams with no or only one magnon-magnon scattering event.
 In this approximation,
we can formally
rewrite  the full Raman intensity in the same form as for
the LF theory:   
\begin{equation}
R (\omega_i, \omega_f) \propto Im \Bigg\{|\bar{M}^{eff}_R|^2 \; 
\frac{I}{1 + I/4S} \Bigg\} \ .
\label{intnew}
\end{equation}
where $I$ is the same as in Eq. (\ref{eff_ver}), and 
$\bar{M}^{eff}_R$ is obtained by 
substituting Eq.(\ref{tripa}) into the diagram in Fig.~\ref{m_eff} and
performing the integration over the intermediate magnon frequency and 
momenta. We then obtain 
\begin{eqnarray}
& & \bar{M}^{eff}_R (\omega_i, \omega_f) \nonumber \\
&& \qquad = i{ 128 J \over N^2} {\sum_{k,q}}^{\prime} 
 {\Big( \displaystyle{\partial \epsilon_k
\over \partial k} e_i \Big) \Big( \displaystyle{\partial \epsilon_{k-q} \over
\partial k} e_f \Big) \Big[ \mu_q \epsilon_{k-q} - \lambda_q \epsilon_k \Big]^2
\over \Big(\omega_i-2E_k+i\Gamma \Big) 
\Big(\omega_f-2E_{k-q}+i\Gamma \Big) } \nonumber  \\
& &  \qquad \times  
{ \tilde{\nu}_q \over 
\Big(\Delta\omega - 2\omega_q +i\delta \Big) 
\Big(\omega_i -\omega_q -E_k-E_{k-q}+i\Gamma \Big)}. 
\label{trip}
\end{eqnarray}
It is essential however that $\bar{M}_R^{eff}$ still possesses a
complex dependence on the external 
incident and outgoing photon frequencies. This in turn 
implies that the full intensity $R (\omega_i, \omega_f)$ is a function
of both frequencies rather than of $\omega_i - \omega_f$ as in the LF theory.
A very similar approach was used by Sch\"{o}nfeld {\it et al.}~\cite{SchKam}.

In our numerical calculations of the full Raman 
intensity we considered all diagrams, i.e., diagrams with 
zero, one or multiple magnon-magnon scattering events. 
The details of this computation are presented in Appendix \ref{full_raman}. 
We found a good qualitative agreement between 
our numerical and analytical results and consider this as a partial
justification for the omission of the 
lowest order diagrams in our analytical considerations.

We now proceed with the discussion of our analytical results, and then present 
our numerical data.

\section{Analytical results}
\label{anal}
In this section, we present the results of our calculations of the Raman
intensity
 near the two resonant frequencies, $\omega^{(1)}_{res}$ and 
$\omega^{(2)}_{res}$. 
We first consider
the case $t^{\prime} =0$ and
then discuss how the intensity changes
 if we break the particle-hole symmetry 
by including a hopping term between
next-nearest neighbors.
Our point of departure is the approximate expression for the Raman intensity,
Eq.(\ref{intnew}),
in terms of the effective Raman vertex, $\bar{M}^{eff}_R$. 
We first study the form of $\bar{M}_R^{eff}$ near 
$\omega^{(2)}_{res}$ and then 
discuss  the form of the vertex near $\omega^{(1)}_{res}$.

\subsection{Resonance at $\omega^{(2)}_{res}$}
\label{om_res2}

\begin{figure} [t]
\begin{center}
\leavevmode
\epsffile{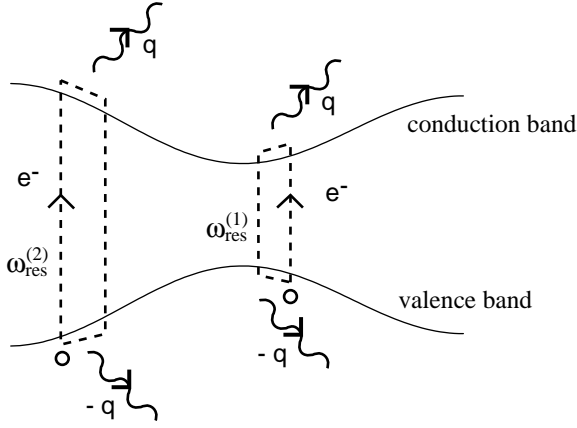}
\end{center}
\caption{Quasiparticles at the bottom of the valence band contribute to the Raman vertex at 
$\omega_{res}^{(2)}$, whereas quasiparticles at the top of the valence band contribute to 
$\omega_{res}^{(1)}$. The dashed and wavy lines represent 
the excited quasiparticles and  
the emitted magnons, respectively.}
\label{bands}
\end{figure}

As we discussed in the introduction, the upper resonance frequency, 
$\omega^{(2)}_{res}$, is close to the maximum possible incident frequency 
$2 \Delta + 8J$ for which the resonance condition can still be satisfied (we assume 
here that $\Delta$ is large, i.e.,  $E_k = \Delta + 
J (\cos k_x + \cos k_y)^2$). The resonance at the highest incident frequency
corresponds to a process in which light causes a transition of 
a quasiparticle from the bottom of the valence band 
to the top of the conduction band (see Fig.~\ref{bands}). 
Since the bottom of the valence band is at ${\bf k}=(0,0)$ and is
not degenerate at the mean-field level, 
it is reasonable to assume that the
corrections to the mean-field fermionic dispersion near the bottom of the band 
will only give rise
to a finite lifetime, but will not introduce any new qualitative features into the
fermionic spectrum. We therefore restrict our considerations to the mean-field form
of the fermionic dispersion and model the fluctuation effects by introducing a
quasiparticle damping $\Gamma$. 

To simplify the presentation, we first neglect the numerator in 
Eq.(\ref{trip}) and focus on the resonance behavior of the effective vertex
due to the vanishing denominator. Near the bottom of the band, we can expand the 
fermionic dispersion and obtain $E_k = \Delta + 4J - 2J k^2$. 
We will assume that near the resonance
the bosonic momentum $q$ is also small such that the expansion
near the bottom of the band holds for both $E_k$ and $E_{k+q}$.
This last assumption will be verified after we perform our calculations.
Further, for small $q$ we can expand $\omega_q$ as $\omega_q = J
\sqrt{2} q$. Substituting the expanded forms of $E_k, E_{k+q}$ and $\omega_q$
into Eq.(\ref{trip}), neglecting the numerator, and 
introducing the dimensionless variables $\lambda_i = (\omega_i -
2\Delta)/2J$ and $\Delta \lambda = \Delta \omega/2J$ we obtain
\begin{eqnarray}
{\bar M}^{eff}_R &\propto&
\int^{\prime} 
\frac{d^2k \;d^2q}{(\Delta \lambda - q \sqrt{2} + i\Gamma)} \nonumber \\
& & \hspace{-1.5cm} \times {1 \over (\lambda_i -4 + 2k^2 +
i\Gamma)(\lambda_i -4 - \Delta \lambda + 2({\vec k}-{\vec q})^2 + i\Gamma) } \nonumber \\
& & \hspace{-2.cm} \times \; {1 \over ( 2\lambda_i -8 - \Delta \lambda + 2k^2 
+2(\vec{k}-\vec{q})^2 + (\Delta \lambda - q \sqrt{2}) + i\Gamma ) } \ . \nonumber \\
\label{trunk}
\end{eqnarray}                             
A similar expansion has been performed by CF.
They however considered only the bare Raman vertex 
(the one without a final state interaction) in which case the Golden Rule
requires the magnons to be on the mass shell, i.e.,
$q = q_0 = \Delta \lambda/\sqrt{2}$. For on-shell magnons, the integration over
$d^2 q$ just yields a factor $2 \pi^2 i q_0$. 
Furthermore, for $q=q_0$,
the last term in the 
denominator in Eq.(\ref{trunk}) is the sum of the second and the third
term, so one can tune $k$ and the angle between ${\bf k}$
and ${\bf q}$ such that all three terms in the denominator of (\ref{trunk})
vanish simultaneously. 
Expanding near this point, CF obtained $M_R \propto (\lambda_i -
\lambda_{res})^{-3/2}$.

We now show that the fully renormalized 
${\bar M}^{eff}_R$ possesses the same
functional form as the bare vertex obtained by CF
since the integration over
$q$ by itself is confined to the vicinity of $q_0$.
 To demonstrate this, we expand near 
the point where the second and
the third term vanish and then integrate over ${\bf k}$ and ${\bf q}$.
The conditions that the two denominators vanish simultaneously 
at a given $q$ are
$ 2k^2=2k^2_0= 4-\lambda_i$ and  $\cos \phi = \cos \phi_0 = (2q^2 - 
\Delta \lambda) / (4 k_0 q)$. We will see below that in order to obtain the
most singular contribution to $\bar{M}^{eff}_R$, one has to
expand around $k_0$ and  $\phi_0$ to linear order in
$x = k-k_0$ and to quadratic order in $y = \phi - \phi_0$.  
Performing this expansion, we obtain
\begin{equation}
\bar{M}^{eff}_R \propto 
\int  \frac{d^2 q \; G(q)}{\Delta \lambda - q \sqrt{2} + i\Gamma} \ ,
\label{Phi}
\end{equation}
where
\begin{eqnarray}
G(q) &=& \int dx \frac{1}{(x + i\Gamma)(k_0 x + 
(\Delta \lambda - q \sqrt{2})/4 + i\Gamma)} \times \nonumber \\
& & \hspace{-1cm} \int dy \Bigg[\frac{1}{C_1 + k_0 q \sin \phi_0 y + k_0 q y^2/2+i\Gamma} 
\nonumber \\
& & \hspace{-.5cm} - \frac{1}{C_2 + k_0 q \sin \phi_0 y + k_0 q y^2/2 + i\Gamma} \Bigg] \ ,
\label{Gq}
\end{eqnarray}                             
and $C_1$ and $C_2$ are given by
\begin{eqnarray*}
C_1 &=& x(k_0 - (2 q^2 - \Delta \lambda)/4 k_0) \ ; \\
C_2 &=& C_1 + k_0 x + (\Delta \lambda - q\sqrt{2})/4  \ .
\end{eqnarray*}
The integration over $y$ can be done explicitly and yields
\begin{eqnarray}
G(q) &=& 2\pi \int \frac{dx}{(x + i\Gamma)(k_0 x + 
(\Delta \lambda - q \sqrt{2})/4 + i\Gamma)} \times \nonumber \\
&&\Bigg[\frac{1}{\sqrt{2 k_0 q C_1 - (k_0 q)^2 \sin^2 \phi_0  + i\Gamma}} \nonumber \\
& &-\frac{1}{\sqrt{2 k_0 q C_2 - (k_0 q)^2 \sin^2 \phi_0 + i \Gamma}} \Bigg] \ .
\label{Gq2}
\end{eqnarray}                             
We immediately see that if $\sin \phi_0$ is finite, 
one can expand the square root in the integrand in Eq.(\ref{Gq2}) in $C_{1,2}$
and find after simple manipulations that
the  Raman vertex is free from singularities.
 To obtain a divergent, resonant contribution to the vertex,
 we therefore have to
set $\sin \phi_0=0$. For any given $q$, this
singles out a line in the $(\lambda_i,\Delta \lambda)$ plane with
\begin{equation}
\lambda_i = \lambda^{(2)}_{res}(q,\Delta \lambda) = 
4 - \frac{q^2}{2} \left(1 - \frac{\Delta \lambda}{2 q^2}\right)^2 \ .
\label{resline}
\end{equation}
Furthermore, we also have to require that the poles and branch cuts in the integrand 
in Eq.(\ref{Gq2}) 
be in different half-planes, since otherwise, the integral over $x$ just 
vanishes.
This requirement is satisfied if $C_{1,2}$, which both are linear functions of
$x$, have negative derivatives $\partial C_{1,2}/\partial x$. 
Near the resonance line $\lambda_i = \lambda^{(2)}_{res}$, we have
\begin{eqnarray*}
C_1 &=& -\frac{x}{4k_0}~\frac{4 q^4 - (\Delta \lambda)^2}{4q^2} \ ;\\
C_2 &=& -\frac{x}{4k_0}~\frac{(2 q^2 - \Delta \lambda) \Delta \lambda}{2q^2} + 
{\Delta \lambda -\sqrt{2}q \over 4} \ .
\end{eqnarray*}
We see that both derivatives are negative if $2 q^2 > \Delta
 \lambda$. If, as we
assume, $q$ is close to $q_0 = \Delta \lambda/\sqrt{2}$, then the 
derivatives are
negative provided $\Delta \lambda >1$. For transferred frequencies near the
two-magnon peak, we have $\Delta \lambda \sim 1.4$, i.e., the above condition is
satisfied. Performing then the integration over $x$, we obtain keeping $\sin
\phi_0$ small but finite
\begin{eqnarray}
G(q) &=& \frac{16 \pi^2}{\Delta \lambda - q \sqrt{2}}~\frac{1}{k_0 q}
 \times \nonumber \\
&&\left[\frac{\sqrt{\sin^2\phi_0 - \frac{\Delta \lambda - q \sqrt{2}}{2 k_0 q}
+ i \Gamma}}{|\sin^2\phi_0 - \frac{\Delta \lambda - q \sqrt{2}}{2 k_0 q}|}
 ~-~\frac{1}{|\sin\phi_0|} \right] \ .
\label{Gq3}
\end{eqnarray}                             
We now express $\sin^2 \phi_0$ in terms of deviation from the critical line as 
\begin{equation}
\sin^2 \phi_0 = \delta - (\Delta \lambda - q \sqrt{2})~\frac{\Delta \lambda
+1}{\Delta \lambda (\Delta \lambda -1)} \ ,
\label{ll}
\end{equation}
where we introduced $\delta = \lambda_i -  \lambda^{(2)}_{res}$.
We now substitute (\ref{Gq3}) and (\ref{ll}) into Eq.(\ref{Phi}) 
and perform the integration over $q$.
Introducing 
$z = (\Delta \lambda - q \sqrt{2})/\delta$ we finally obtain
\begin{equation}
\bar{M}^{eff}_R \propto ~\frac{Z(\Delta \lambda)}{\Delta \lambda }~\frac{1}{\delta^{3/2}} \ ,
\label{phi1}
\end{equation}
where
\begin{eqnarray*}
Z (\Delta \lambda) &=& 2 (1 + \Delta \lambda)~\int_{-\infty}^1 {d z \over (z-i\Gamma)^2} \\
&& \hspace{-0.5cm} \times \Bigg\{ {1 \over \sqrt{1-z}}
 -{1 \over \sqrt{1-z + \frac{z-i\Gamma}{2 (1 + \Delta \lambda)}}}
\Bigg\} 
\end{eqnarray*}
is a very smooth (nearly a constant) function of the momentum transfer.
Clearly, the integral over $z$ comes from $z=O(1)$ which in turn implies that
the actual integral over $q$ is confined to the region $\Delta \lambda - q \sqrt{2} =
O(\delta)$. The fact that
$q$ is confined to the  vicinity of its on-shell value implies
that one can set $q = \Delta \lambda/\sqrt{2}$ in Eq.(\ref{resline})
which yields
\begin{equation}
\lambda^{(2)}_{res}(\Delta \lambda) = 4 - \frac{(\Delta \lambda -1)^2}{4} \ .
\label{rl}
\end{equation}
For $\Delta \lambda=1.4$, which, we remind, corresponds to the position of the two 
magnon peak, we obtain 
$\lambda^{(2)}_{res}=3.93$.
Also, for this $\Delta \lambda$ we have $q \approx 1$. This $q$ is not small enough to fully
justify our expansion in the magnon momentum. However, for $q_x = q_y$, the approximation
of $\omega_q$ by a linear term is incorrect only by 8\%. 

We see that $\bar{M}^{eff}_R$ diverges as $\delta^{-3/2}$ when
the incident frequency approaches the resonance value $\lambda^{(2)}_{res}$. 
The same result has been obtained by CF who neglected the final state interaction.
In this respect, our result shows 
that the final state  interaction
does not destroy the triple resonance which occurs along the same line as in the
absence of the magnon-magnon scattering.

We now consider the effect of the  numerator in  $\bar{M}_R^{eff}$.
In principle, the numerator is finite at $\lambda_i = \lambda^{(2)}_{res}$ so that
the $\delta^{-3/2}$ behavior should survive close to the resonance line. In practice, however,
for transferred frequencies around $\Delta \lambda =1.4$, 
the resonance value of $\lambda_i$ is close to the maximum 
possible resonance value of $4$. At this
maximum value of $\lambda_i$ the matrix element for
the interaction between light and the fermionic current vanishes and the numerator turns to zero.
A simple analysis 
similar to the one performed by CF shows that the numerator vanishes 
as $4-\lambda_i$ for $\lambda_i \rightarrow 4$. The full 
$\bar{M}^{eff}_R$ then behaves as  
\begin{equation}
\bar{M}^{eff}_R = A (\Delta \lambda)~\frac{4 - \lambda_i}
{|\lambda_i - \lambda^{(2)}_{res}|^{3/2}} \ .
\label{res1}
\end{equation}
Restoring all numerical factors in the expression for the 
Raman vertex, we find that $A(\Delta \lambda)$ has the form
\begin{equation}
A (\Delta \lambda) = \frac{J}{\pi}~\left(\frac{t}{2J}\right)^4 (1 + \Delta \lambda)~
Z(\Delta \lambda)\ .
\end{equation}
At some distance
away from $\lambda^{(2)}_{res}$, the difference between $4 - \lambda_i$ and 
$\lambda^{(2)}_{res} - \lambda_i$
is irrelevant and $\bar{M}^{eff}_R$ can be approximated by 
\begin{equation}
\bar{M}^{eff}_R = 
\frac{A (\Delta \lambda)}{|\lambda^{(2)}_{res} - \lambda_i|^{1/2}} \ .
\end{equation}
This is the final expression for the effective Raman vertex. Substituting this
expression into  Eq.(\ref{intnew}), one obtains for the Raman intensity
\begin{equation}
R(\lambda_i, \Delta \lambda) \propto \frac{|A (\Delta \lambda)|^2}{|\lambda^{(2)}_{res} -  
\lambda_i|}~~\frac{I}{1 + I/4S} \ ,
\label{newR}
\end{equation}
where $I$ is given in Eq.(\ref{eff_ver}).
This result for $R$ is in fact very similar to what CF conjectured: the full intensity
is a product of two terms, one, $I/(1+I/4S)$,  depends on $\Delta \lambda$ and has the
same form as in the LF theory, while the other, $|\lambda_i - \lambda^{(2)}_{res}|^{-1}$, 
in essence reflects the enhancement of the bare Raman vertex. This last term depends on
$\lambda_i$ but also on the transferred frequency through $\lambda^{(2)}_{res}$. 
There is however an extra feature: the interplay
between triple resonance and magnon-magnon interaction gives rise to an extra
dependence of the intensity on $\Delta \lambda$, through the factor $A$. However, we
found that $A$ is a smooth function of momentum transfer, so this extra dependence 
is not that relevant. 
This is particularly true near $\lambda^{(2)}_{res}$ where the two-magnon peak is rather narrow
so one probes $A$ only in a narrow range of $\Delta \lambda$ around $1.4$.
We already mentioned in the introduction that the inverse linear dependence of the TMPH
near $\omega^{(2)}_{res}$ was first experimentally observed
by Blumberg {\it et al.} for $YBa_2Cu_3O_{6.1}$~\cite{BluAbb} and later by R\"{u}bhausen {\it et 
al.} for $PrBa_2Cu_3O_7$~\cite{RubDie}. This is fully consistent with our result. We, however, 
performed our calculations only 
in the vicinity of the resonance and therefore cannot provide a theoretical estimate for 
the width of the region where the inverse linear behavior holds.
The data for $Sr_2CuO_2Cl_2$ are a bit less conclusive because
in  this material, $\omega_{res}^{(2)}$ is  larger than the
largest experimentally accessible $\omega_i$,
and one cannot unambiguously conclude from the data that
the Raman intensity follows an inverse linear behavior in a wide range of frequencies.

Besides the behavior of the TMPH near $\lambda^{(2)}_{res}$, Eq.(\ref{newR}) also 
describes the form of the two-magnon profile at a given $\lambda_i$.
The conventional factor $I/(1+I/4S)$ produces a symmetric
peak in the intensity at $\Delta \lambda =1.4$. The other two factors which 
contribute to the peak lineshape
are the dependencies on $\Delta \lambda$ in the overall factor $A$ and in  
$\lambda^{(2)}_{res}$. Near the resonance value $\lambda_i =3.93$,
we found that the two extra contributions to the lineshape cancel each other and
the resulting two-magnon profile is chiefly given by $I/(1+I/4S)$ and therefore
symmetric. For incident frequencies smaller than 
$\lambda_i=3.93$, the triple resonance in the Raman vertex occurs at transferred
frequencies larger than $\Delta \lambda =1.4$. This obviously gives rise to an asymmetry
of the two-magnon lineshape with a larger intensity at higher frequencies.  
The evolution from an asymmetric to a symmetric form of the two magnon profile when
$\lambda_i$ approaches the resonance value from below is consistent with the experimental
data. We will also demonstrate this effect when we discuss our numerical results.

\subsection{Resonance near $\omega^{(1)}_{res}$}
\label{om_res1}

The triple resonance theory of CF  predicts that the TMPH measured as a
function of $\omega_i = 2\Delta + 2 J \lambda_i$ exhibits a second maximum  
at a relatively small frequency $\omega^{(1)}_{res}=2\Delta +3J$ (see Fig.~\ref{trip_res}). 
For this low frequency resonance  CF found a much smaller
range of magnon momenta for which the resonance conditions are satisfied and therefore
concluded that this resonance should be weaker than the one at
$\omega^{(2)}_{res}$. We now discuss this issue in more detail and will show
that while the divergent term at $\omega^{(1)}_{res}$ is almost completely 
suppressed, the subleading terms are larger than at $\omega^{(2)}_{res}$.
 
In the previous
subsection,  we expanded the fermionic  dispersion
around the bottom of the valence band since $\omega^{(2)}_{res}$ is close to
the maximum possible incident frequency for the triple resonance.
Here, on the contrary, we will make use of the
fact that the low-frequency resonance occurs at $\omega^{(1)}_{res} \approx
 2\Delta + 3J$ which is not far from 
the minimum incident frequency ($=2\Delta+2J$) for which a triple resonance is possible. 
Accordingly, we will study the
behavior of the Raman intensity by expanding the fermionic dispersion
upto quadratic order around the top of the valence band.  As in the previous 
subsection, we will also expand the magnon dispersion to linear order in the 
momenta.
We will see that in this approximation the Raman vertex is free from actual divergencies.

A peculiarity associated with the expansion of the fermionic dispersion near the 
top of the valence band is that the position of the band maximum 
is degenerate at the mean-field level. Numerous
analytical and numerical calculations, however, have demonstrated that
this degeneracy is an artifact of the mean-field treatment 
\cite{Scal,ChuMus,Duff,Hanke,Dag} -  the actual
fermionic dispersion possesses a maximum at $(\pi/2, \pi/2)$ 
and symmetry related points. 
Recent photoemission experiments on 
$Sr_2CuO_2Cl_2$ confirmed this result and in addition have shown that  
the dispersion near the
top of the valence band is nearly isotropic around $(\pi/2,\pi/2)$ \cite{Wells,LaR}. 
At the moment it is still a topic of controversy, whether one needs a substantially large 
next-nearest-neighbor hopping to explain an
almost isotropic dispersion, or whether 
it is a property of the nearest-neighbor model in the strong coupling limit, as was first 
suggested by Laughlin \cite{Lau}.
The set of parameters suitable to $Sr_2CuO_2Cl_2$, namely $J/t \sim 0.4$,
corresponds to an intermediate coupling regime in which case the
next-nearest-neighbor exchange is probably needed to account for the isotropy of the
spectrum \cite{Dirk1}. This second-neighbor hopping breaks the particle-hole symmetry and 
causes substantial
complications for the calculation of the Raman vertex since Eq.(\ref{trip}) is no longer valid.
At the same time, we know that the degeneracy along the 
magnetic Brillouin zone boundary is
already lifted in the nearest-neighbor model due to self-energy corrections. One can therefore 
argue that the inclusion of $t^{\prime}$ only 
aids in fitting the ratio of the effective masses but does {\it not} introduce any new 
qualitative features.
 To avoid unnecessary complications, we will assume 
that the particle-hole symmetry is preserved, and that the experimentally measured
nearly isotropic quasi-particle dispersion around the 
top of the valence band is due to strong self-energy corrections. In
other words, we will 
still use  Eq.(\ref{trip}) and will also assume that near the top of
the valence band we can expand the fermionic dispersion as 
$E_k = \Delta + J {k}^2$, where $k$ 
measures the deviation from $(\pi/2,\pi/2)$.
Substituting the  expansion for $E_k$ into Eq.(\ref{trip}) and neglecting the
numerator which is finite near the low-frequency resonance, 
we obtain 
\begin{eqnarray}
\bar{M}^{eff}_R &\propto& 
\int^{\prime} \frac{d^2 k \; d^2 q}{ \Delta \lambda - q \sqrt{2} + i\Gamma } \nonumber \\
& & \hspace{-1cm} \times {1 \over (\lambda_i  - k^2 +i\Gamma)(\lambda_i - \Delta \lambda - 
({\vec k}-{\vec q})^2 + i\Gamma)} \nonumber \\
& & \hspace{-1.5cm} \times {1 \over (2\lambda_i - \Delta \lambda - k^2 - ({\vec k}-{\vec q})^2 + 
(\Delta \lambda - q \sqrt{2}) + i\Gamma)} \ .
\label{trunk2}
\end{eqnarray}
To study whether the effective Raman vertex diverges  at some particular
$\lambda_i$, we perform the same analysis as in the previous section, i.e.,
expand near a particular fermionic momentum, ${\bf k}_0$ where
the second and third terms in the denominator vanish 
simultaneously at a given ${\bf q}$.  For $k_0$ and $\phi_0$,
which is the angle between ${\bf k}_0$ and ${\bf q}$ we find
$k^2_0 = \lambda_i$ and 
$\cos \phi_0 =(q^2 + \Delta \lambda)/(2 k_0 q)$. 
Expanding, as before, around $k_0$ and  $\phi_0$
to linear order in $x = k-k_0$ and to quadratic order in $y = \phi - \phi_0$, 
we obtain
\begin{equation}
\bar{M}^{eff}_R \propto 
\int d^2 q \frac{ \tilde{G}(q) }{ \Delta \lambda - q \sqrt{2} + i\Gamma } \ ,
\label{Phi2}
\end{equation}
where 
\begin{eqnarray}
{\tilde G}(q) &=& \int dx \frac{1}{(x + i\Gamma)(k_0 x + 
(\Delta \lambda - q \sqrt{2})/4 + i\Gamma)} \times \nonumber \\
& & \times \int dy \Bigg[\frac{1}{{\tilde C}_1 + k_0 q \sin \phi_0 y - k_0 q y^2/2 +i\Gamma} 
\nonumber \\
& & -\frac{1}{{\tilde C}_2 + k_0 q \sin \phi_0 y - k_0 q y^2/2 + i \Gamma} \Bigg] \ ,
\label{Gq22}
\end{eqnarray}                             
and  ${\tilde C}_1$ and ${\tilde C}_2$ are given by
\begin{eqnarray*}
\tilde{C}_1 &=& x(k_0 - (q^2 + \Delta \lambda)/2 k_0) \ ; \\
\tilde{C}_2 &=& C_1 + k_0 x + (\Delta \lambda - q\sqrt{2})/4 \ .
\end{eqnarray*}
The integration over $y$ can  again be performed 
explicitly, and we obtain 
\begin{eqnarray}
{\tilde G}(q) &=& -2\pi i~ \int dx \frac{1}{(x + i\Gamma)(k_0 x + 
(\Delta \lambda - q \sqrt{2})/4 + i\Gamma)} \times \nonumber \\
& & \Bigg[\frac{1}{\sqrt{2 k_0 q {\tilde C}_1 + (k_0 q)^2 \sin^2 \phi_0 
 + i\Gamma}} \nonumber \\
& &-\frac{1}{\sqrt{2 k_0 q {\tilde C}_2 + (k_0 q)^2 \sin^2 \phi_0 + i \Gamma}} \Bigg]  \ .
\label{Gq222}
\end{eqnarray}
This expression is similar to Eq.(\ref{Gq2}) and we again find that  
if $\sin \phi_0$ is finite, one can expand the square root
and obtain that ${\tilde G} (q)$ is free from singularities. The expansion
does not work, however, if $\sin \phi_0 = 0$.  In this case, a
power counting argument indicates that the Raman vertex 
may diverge. The condition $\sin \phi_0 = 0$
singles out a line in the $(\lambda_i,\Delta \lambda)$ plane with
\begin{equation}
\lambda_i = \lambda^{(1)}_{res}(q,\Delta \lambda) = 
 \frac{q^2}{4} \left(1 +\frac{\Delta \lambda}{q^2}\right)^2 \ .
\label{resline1}
\end{equation}
The very existence of the critical line along which the Raman vertex
diverges by the power
counting argument, however, does not 
guarantee that the divergence is actually genuine.
Indeed, the arguments displayed in the previous subsection show that the
vertex diverges only if the poles and the branch cuts in Eq.(\ref{Gq222}) are
located in different half-planes. This is the case if the derivatives over $x$
of ${\tilde C}_{1,2}$ are negative. However, near $\lambda^{(1)}_{res}$ we have
\begin{eqnarray*}
\tilde{C}_1 &=& \frac{x}{2k_0}~\frac{2 + \Delta \lambda}{4} (2 - \Delta \lambda) \ ;\\
\tilde{C}_2 &=& C_1 + x k_0 + { \Delta \lambda - q\sqrt{2} \over 4} \ , 
\end{eqnarray*}
Since $(2 - \Delta \lambda)$ is always positive, the derivatives are
clearly positive in which case 
the divergent contribution vanishes after the integration over $x$.
We see therefore that the triple resonance does not yield 
a divergence in the Raman vertex at $\lambda_i = \lambda^{(1)}_{res}$, contrary to
what we found near $\lambda^{(2)}_{res}$.\\ 
We then studied the form of the Raman vertex in more detail and found that
the absence of a divergence is a result of the restriction to a
quadratic dispersion around the top of the band. Expanding further in $k$
and redoing the calculations, we obtained a
divergence in $M_{R}$ resulting from the integration over a small region of 
fermionic momenta. A similar result was also obtained by CF who used a somewhat
different technique. 
However, the phase factor associated with the divergent
contribution to $M_R$ is very small, and the divergence is already eliminated 
by a small fermionic damping.
 
So far, we have found that the Raman vertex exhibits a regular behavior around
$\lambda^{(1)}_{res}$. Experimentally, however, the TMPH clearly displays
a second maximum at $\lambda_i \approx 1.5$.
We will now show that
this maximum can in fact be described within the triple resonance theory
since the Raman vertex turns out to be
very strongly enhanced near $\lambda^{(1)}_{res}$. 
To demonstrate this, 
it is not sufficient to expand near particular values of $k_0$ and $\phi_0$,
and we thus need to study the full form of the Raman vertex.
The full form of $M_R$ near $\lambda^{(2)}_{res}$ was first obtained by CF and
we now have to perform the same analysis near $\lambda^{(1)}_{res}$.
There is, however, a subtlety related to these calculations. 
CF had to assume that the magnons are on-shell, i.e., $q=q_0 = \Delta \lambda/2$,
since a full analytical consideration is not possible without this last assumption. 
In the previous subsection, we found that the divergent piece of the vertex
is intrinsically confined to a narrow range around $q_0$, and
the results with and without the
restriction to only on-shell magnons are roughly the same.
Near $\lambda^{(1)}_{res}$, the situation is less rigorous since the divergent
piece is absent. At the same time, it still looks
reasonable to estimate the value of the vertex near $\lambda^{(1)}_{res}$
by just restricting with
on-shell magnons. Doing this and following the computational steps outlined by
CF, we obtain  after some lengthy calculations 
\begin{eqnarray}
\bar{M}^{eff}_R &\propto& 
{D_1 \over |{\tilde \delta}|^{3/2}} 
\left[ \log {1 - \sqrt{2 \tilde{\delta} } -i\Gamma \over 
1 + \sqrt{2 \tilde{\delta} } + i\Gamma} - \log {1 - 
\frac{2\sqrt{2 \tilde{\delta} } }{2 + \Delta \lambda} - i\Gamma  \over 1 + 
\frac{2\sqrt{2 \tilde{\delta}}}{2 + \Delta \lambda} + i\Gamma} \right] \nonumber \\
& & \hspace{-1cm} + \frac{D_2}{|{\tilde \delta}|^{3/2}}
 \left[\log  {1 + \sqrt{2 \tilde{\delta} } +i\Gamma \over
1 - \sqrt{2 \tilde{\delta} } - i\Gamma} - \log  {1 + 
\frac{2\sqrt{2 \tilde{\delta} } } {2 -\Delta \lambda} + i\Gamma  \over 1 -
\frac{2\sqrt{2 \tilde{\delta} } } {2 -\Delta \lambda} - i\Gamma } \right] \ , \nonumber \\
\label{uff}
\end{eqnarray}
with $D_1 = (2+ \Delta \lambda)/(\sqrt{2} (\Delta \lambda)^2)$, 
$D_2 = (2- \Delta \lambda)/(\sqrt{2} (\Delta \lambda)^2)$, and 
${\tilde \delta} = \lambda^{(1)}_{res} -\lambda_i$
where $\lambda^{(1)}_{res}$ is given by Eq.(\ref{resline1}) with $q$ substituted by
$q_0$:
\begin{equation}
\lambda^{(1)}_{res}(\Delta \lambda) = \frac{1}{2}~\left(1 + \frac{\Delta \lambda}{2}\right)^2 
\ .  
\label{rl1}
\end{equation} 
Though both terms in
Eq.(\ref{uff}) contain a term $|\tilde{\delta}|^{-3/2}$, the combination 
of logarithms vanishes when $\tilde{\delta}$ approaches zero. 
Moreover, expanding in $\tilde{\delta}$, we find that the terms of $O(\tilde
{\delta}^{1/2})$ also cancel each other.
Consequently,  there is not even a weak triple resonance
at $\lambda^{(1)}_{res}$, and the Raman vertex
turns out to be a regular  function 
of $\lambda_i$ in the immediate vicinity of the would-be resonance line 
$\lambda^{(1)}_{res}$. This indeed agrees with our expansion near
$k_0$ and $\phi_0$. At the same time, it follows from 
Eq.(\ref{uff}) that at frequencies only slightly smaller than 
$\lambda^{(1)}_{res}$, namely at 
$(2 -\Delta \lambda)/2 < \sqrt{2 {\tilde \delta}}$, one of the
logarithms contains an extra $i\pi$ factor associated with the branch cut.
Due to this extra factor, $\bar{M}^{eff}_R$ in fact scales as 
$|\tilde{\delta}|^{-3/2}$, i.e., the vertex possesses the
same functional dependence on the incident frequency as if the triple resonance
at $\lambda^{(1)}_{res}$ were actually present.
Near the two-magnon peak we have $\Delta \lambda = 1.4$.
In this case the singular behavior actually starts
very close to $\lambda^{(1)}_{res}$, namely at 
$ \lambda^{(1)}_{res} - \lambda_i \sim 0.045$. 
This singular behavior exists, with decreasing amplitude, upto
$2 \tilde{\delta} = 1$, or $\lambda^{(1)}_{res} -\lambda_i \sim 0.5$, though at such high deviations from
$\lambda^{(1)}_{res}$ the regular and singular parts of 
${\tilde G} (q)$ are of the same order.  
We see therefore that despite the absence of a true divergence at
$\lambda^{(1)}_{res}$, the TMPH still possesses a maximum very close to it.
Moreover, for $\Delta \lambda =1.4$, we have
$\lambda^{(1)}_{res}=\lambda^{(1)}_{res} \sim 
1.45$, i.e., $\omega^{(1)}_{res} \sim  2\Delta + 3J$ which is in good
agreement with the experimentally observed location 
of the low-frequency peak in the TMPH.\\            
For experimental comparisons, it is essential that the enhancement due to the branch cut 
in $M_R$ is asymmetric - it exists for $\lambda_i < \lambda^{(1)}_{res}$ but not
for $\lambda_i > \lambda^{(1)}_{res}$. This should obviously yield an asymmetric form 
of the TMPH near $\lambda^{(1)}_{res}$ - the intensity should increase continuously
as one approaches $\lambda^{(1)}_{res}$ from below and drop down rather fast when
$\lambda_i$ exceeds $\lambda^{(1)}_{res}$. In addition we should also obtain an
asymmetry of the two-magnon lineshape at exactly $\lambda_i = \lambda^{(1)}_{res}$ 
with a higher intensity at larger frequencies.
Indeed, for  $\Delta \lambda >1.4$ we have $\lambda^{(1)}_{res} >
\lambda_i$, and the Raman vertex is strongly enhanced. No such effect, however, exists 
for $\Delta \lambda <1.4$. Both above mentioned anisotropies are consistent with the
experimental data.
 
Finally, consider the numerator in the Raman vertex. Near
$\lambda^{(2)}_{res}$, the numerator was small due to the proximity to the bottom
of the band, and effectively reduced 
the divergence of the Raman vertex to $\delta^{-1/2}$ instead of $\delta^{-3/2}$. 
The lower resonance frequency, 
$\lambda^{(1)}_{res}$, however, is rather far from the bottom of the band
so that the numerator does not possess any smallness. As a result, 
the strong enhancement of the Raman vertex as one approaches 
$\lambda_{res}^{(1)}$ from below 
turns out to be comparable, and for some values of
parameters even larger than the intensity near the high-frequency resonance.
We will explicitly demonstrate this feature in our numerical results
in Sec.~\ref{tmph}.

\subsection{Raman intensity at finite $t^{\prime}$}
\label{om_res2a}

We now consider how the inclusion of a next-nearest neighbor 
hopping modifies the resonant behavior of the Raman vertex. 
We already discussed above that a nonzero $t^{\prime}$ breaks
the particle-hole symmetry. In this case, the expression 
for the Raman vertex is more complex than Eq.(\ref{tripa}) and has the form
\begin{eqnarray}
& & M^{\prime}_R= -{ 4i \over N} {\sum_k}^{\prime}  {\Big( \displaystyle{\partial 
\tilde{\epsilon}_k
\over \partial k} e_i \Big) \Big( \displaystyle{\partial \tilde{\epsilon}_{k-q} \over
\partial k} e_f \Big) \Big[ \mu_q \epsilon_{k-q} - \lambda_q \epsilon_k \Big]^2
\over \Big(\omega_i-2E_k+i\Gamma \Big) \Big(\omega_f-2E_{k-q}+i\Gamma \Big) } \nonumber  \\
& &  \quad \times  \Bigg\{ {1 \over \Big(\omega_i- \omega + E_k^v
-E_{k-q}^c +i\Gamma \Big)} \nonumber \\
& & \qquad \ \ +  
{1 \over \Big(\omega_f +\omega -+E_{k-q}^v-E_k^c+i\Gamma \Big)}  \Bigg\} \ ,
\label{tripa_tp}
\end{eqnarray} 
where $\epsilon_k$ and $E_k$ are defined as before and 
\begin{eqnarray*}
\tilde{\epsilon}_k &=& -4t\nu_k-4t^{\prime} cos(k_x)cos(k_y), \\ 
E_k^{c,v}&=&\pm \sqrt{\Delta^2+\epsilon_k^2}-4t^{\prime} cos(k_x)cos(k_y).
\end{eqnarray*}
Here $E_k^{c,v}$ describes 
the energy dispersion of the conduction and valence bands, 
respectively. 

One can easily see that at finite $t^{\prime}$,
only two out of the three terms in the denominator 
in Eq.(\ref{trip_tp}) can vanish simultaneously. A nonzero $t^\prime$ thus effectively 
transforms the triple resonance into a set of double resonances.
Obviously, there exist five combinations 
of terms for which the denominator can vanish.
One of these, namely the one with $\omega_i-2E_k=0$ and $\omega_f-2E_{k-q}=0$, 
yields a resonance in exactly the same region of the 
$( \omega_i, \Delta \omega)$ plane 
where the resonance occurs without $t^{\prime}$. We will show that of the remaining 
four combinations only 
two are truly divergent in the vicinity of $\omega^{(2)}_{res}$.

In order to obtain some analytical results we again have to calculate the
effective Raman vertex which now has a more complex form:
\begin{eqnarray}
& & \bar{M}^{eff}_R= i{ 64 J \over N^2} {\sum_k}^{\prime} {\sum_q}^{\prime}
 {\Big( \displaystyle{\partial \tilde{\epsilon}_k
\over \partial k} e_i \Big) \Big( \displaystyle{\partial \tilde{\epsilon}_{k-q} \over
\partial k} e_f \Big) \Big[ \mu_q \epsilon_{k-q} - \lambda_q \epsilon_k \Big]^2
\over \Big(\omega_i-2E_k+i\Gamma \Big) 
\Big(\omega_f-2E_{k-q}+i\Gamma \Big) } \nonumber  \\
& &  \qquad \times  
{ 1\over \Big(\Delta\omega - 2\omega_q +i\Gamma \Big)}~
\Bigg[{ 1\over 
\Big(\omega_i -\omega_q +E^{v}_k-E^{c}_{k-q}+i\Gamma \Big)} \nonumber \\
& & \hspace{1cm} +
{ 1\over 
\Big(\omega_i -\omega_q -E^{c}_k+E^{v}_{k-q}+i\Gamma \Big)} \Bigg] \ .
\label{trip_tp}
\end{eqnarray}                                              
We first observe that while
the energy dispersion of the quasiparticles depends linearly on $t^{\prime}$,
the magnon dispersion and the magnon-magnon scattering vertex $V(k,q)$
depend only on $(t^{\prime})^2$. Since $(t^{\prime}/t)^2
<0.25$ (otherwise, the antiferromagnetic state is unstable), we will just 
neglect $t^{\prime}$ in $\omega_q$ and $V(k,q)$.

Consider first the situation near
 $\omega_{res}^{(2)}$ when there is a true
resonance in $\bar{M}^{eff}_R$. 
Performing now the same manipulations as before, i.e., 
expanding near the bottom of the  
band and neglecting the numerator in Eq.(\ref{trip_tp}), we obtain $\bar{M}^{eff}_R
= (\bar{M}^{eff}_R (a) + \bar{M}^{eff}_R (-a))/2$, where $a = t^{\prime}/J$ and 
\begin{equation}
\bar{M}^{eff}_R (a) \propto 
\int d^2 q~\frac{G_a (q)}{(\Delta \lambda - q \sqrt{2} + i\Gamma)}
\label{trunk3}
\end{equation}
with
\begin{eqnarray}
G_a (q) &=& \int d^2 k \ (\lambda_i -4 + 2k^2 +
i\Gamma)^{-1} \nonumber \\
& & \quad  \times (\lambda_i -4 - \Delta \lambda + 2({\vec k}-{\vec q})^2 + i
\Gamma)^{-1} \nonumber \\
& & \hspace{-1cm} \times \Big[ 2\lambda_i -8 - \Delta \lambda + 2k^2 (1+a)
 +2({\vec k}-{\vec q})^2 (1-a) \nonumber \\
 & & + (\Delta \lambda - q \sqrt{2}) + i\Gamma \Big]^{-1} \ .
\end{eqnarray}                             
Expanding near the point where the first two terms in the 
denominator in $G_a (q)$
vanish and integrating over the deviations from the resonance values, 
we obtain for $G_a(q)$ 
\begin{eqnarray}
G_a (q) &=& 2\pi \int dx \frac{1}{(x + i\Gamma)(k_0 x (1+a)/(1-a)+ 
z + i\Gamma)}  \nonumber \\
&& \times \Bigg[\frac{1}{\sqrt{2 k_0 q C^a_1 \cos \phi_0 
- (k_0 q)^2 \sin^2 \phi_0 + i\Gamma}} \nonumber \\
& &- \frac{1}{\sqrt{2 k_0 q C^a_2 \cos \phi_0 
- (k_0 q)^2 \sin^2 \phi_0 + i \Gamma}} \Bigg] \ .
\label{Gq33}
\end{eqnarray}                             
Here $k_0 = [(4-\lambda_i)/2]^{1/2}$ and $\cos \phi_0 = (2q^2 - \Delta \lambda)/4k_0 q$
are the same as for the resonance with $t^{\prime}=0$, 
$z = (\Delta \lambda (1+a) - q \sqrt{2})/4(1-a)$, and $C^a_{1,2}$ are
given by
\begin{eqnarray*}
C^a_1 &=& x(k_0 - q\cos \phi_0) \ ; \\
C^a_2 &=& x(2 k_0/(1+a) - q \cos \phi_0 z) \ .
\end{eqnarray*}
One can easily verify that  the derivatives of $C^a_{1,2}$ are negative 
in which case the poles and branch cuts in Eq.(\ref{Gq33}) are located in different
half-planes. This implies that the integral over $x$ is finite. 
Performing the explicit integration over $x$,
we obtain after some simple manipulations
\begin{eqnarray}
G_a (q) &=& \frac{4 \pi^2}{z}~\frac{1}{k_0 q}
 \times \nonumber \\
&&\left[\frac{\sqrt{\sin^2\phi_0 - \frac{2z \cos \phi_0}{k_0 q}
+ i \Gamma}}{|\sin^2\phi_0 - \frac{2z \cos \phi_0}{k_0 q}|}
 ~-~\frac{1}{|\sin\phi_0|} \right] \ .
\label{Gq34}
\end{eqnarray}                             
Now we are left with the integral over $q$ in Eq.(\ref{trunk3}). 
In Sec.~\ref{om_res2}, the $q$ integration was confined to
a narrow region around $q_0 = \Delta \lambda / \sqrt{2} $ and
 yielded $\bar{M}^{eff}_R \propto
|\delta|^{-3/2}$ where
 $\delta= \lambda_i - \lambda^{(2)}_{res}$. At finite $t^{\prime}$,
an analysis of Eq.(\ref{Gq34}) shows that 
there exist two regions in momentum space which yield singular 
contributions to ${\bar M}^{eff}_R (a)$. The first region is still
 the vicinity of $q = q_0 = \Delta \lambda / \sqrt{2}$. However,
 since $z$ is finite for $q=q_0$, this region yields a weaker,
$\delta^{-1/2}$ singularity in ${\bar M}^{eff}_R (a)$.
In practice, everywhere except for the immediate vicinity of the resonance, 
this divergence is fully compensated by the numerator in $\bar{M}^{eff}_R (a)$ which
vanishes linearly as $\lambda_i$  approaches $\lambda_i^{max}=4$ which is very close to
$\lambda^{(2)}_{res}$. 
We checked that the same functional behavior also holds for $M^{eff}_R (-a)$.
 
The second  singular contribution to $\bar{M}^{eff}_R (a)$ comes from the $q$
integration over the region where $z$ is nearly zero. The conditions $z=0$ and
$\sin \phi_0 =0$  specify a line in the ($\lambda_i, \Delta \lambda$) 
plane with
\begin{equation}
\lambda_i = \lambda^{(2)}_{res} (a,\Delta \lambda) 
= 4 - \frac{(1 - \Delta \lambda (1+a)^2)^2}{4(1+a)^2} \ .
\label{res_2a}
\end{equation}
We found that near this line, the Raman vertex also 
diverges as $\delta^{-1/2}$ where 
$\delta$ now measures
the deviation from $\lambda^{(2)}_{res} (a)$. 
This square root divergence near $z=0$ also holds for 
$\bar{M}^{eff}_R (-a)$ for which the resonance incident frequency is 
given by Eq.(\ref{res_2a}) with $a$ replaced by $-a$. 
We see therefore that a nonzero $t^{\prime}$ splits the strong resonance
at $\lambda^{(2)}_{res}$ with a $\delta^{-3/2}$ singularity 
into three weaker resonances with $\delta^{-1/2}$ singularities. 
One of these weaker resonances still occurs 
at $\lambda^{(2)}_{res}$ while the two new resonances occur at
$\lambda^{(2)}_{res} (\pm a)$.
For $a=0$, the three resonance lines coincide 
and we recover the result of Sec~\ref{om_res2}.

For $a =-0.5$, which is relevant to the cuprates, 
and $\Delta \lambda = 1.4$
 we have $\lambda^{(2)}_{res} = 3.93$ while 
$\lambda^{(2)}_{res} (a) \approx 3.49$ and $\lambda^{(2)}_{res} (-a) \approx
3.58$. We see that the new resonance frequencies are further away
from $\lambda_i^{max}$ than $\lambda^{(2)}_{res}$ and therefore should be 
less effected by the smallness of the numerator in $\bar{M}^{eff}_R$. 
Naively, this should make the 
new resonances stronger than the one at $\lambda^{(2)}_{res}$. However, 
we found that the overall numerical factor
is larger near $\lambda^{(2)}_{res}$ than near  the two new resonance lines. 
In this situation, $t^{\prime}$ just reduces and broadens the
peak at $\lambda^{(2)}_{res}$ without actually producing comparable peaks at
the two new resonance frequencies.
Our numerical findings in Sec.~\ref{tprime} 
are fully consistent with this result.

Finally, we shortly discuss the effect of $t^{\prime}$ on the low-frequency resonance.
We found that the inclusion of $t^{\prime}$ 
shifts the frequency range for the enhancement due to the branch cut but does 
not introduce any new physics near $\lambda^{(1)}_{res}$.  
We again obtained 
that there is no real divergence of the Raman vertex in this frequency range
but  that slightly below $\lambda^{(1)}_{res}$, the vertex acquires a branch cut
enhancement which mimics the resonance behavior.
The calculations near $\lambda^{(1)}_{res}$ 
are, however, rather involved, and we did not succeed in 
fully solving the problem analytically.
We will discuss our numerical results near $\lambda^{(1)}_{res}$  
in Sec.~\ref{tprime}. 

\section{Numerical results} 
\label{numerics}
In the following subsections we 
will present our numerical results for the Raman lineshape 
and the TMPH. In Secs.~\ref{shape} and \ref{tmph} we first consider a system with 
particle-hole symmetry. In Sec.~\ref{tprime} we study how the form of the TMPH is 
modified due to a finite 
next-nearest hopping term $t^{\prime}$ which breaks the particle-hole 
symmetry. Finally, in Sec.~\ref{vertex} we discuss how the renormalization of the
interaction  between light and quasiparticles due to vertex corrections affects the form of the 
TMPH. We  summarize all relevant formulae for the numerical 
computation of the Raman intensity with 
final state interaction in Appendix \ref{full_raman}.

Before we proceed, we want to point out the differences in our numerical and 
analytical considerations for $\omega_i \approx \omega_{res}^{(1)}$. For our 
numerical calculations we use the mean-field form of the fermionic excitation spectrum, 
which in the case $t^\prime=0$ is degenerate along the boundary of the magnetic Brillouin 
zone. This particular form of the dispersion yields, besides an enhancement due
to a branch cut, also a real divergence of $M_R$
at $\omega_{res}^{(1)}$, though with a small overall factor.
In our analytical calculations in Sec.~\ref{om_res1} we 
replaced this mean-field form by a quadratic dispersion around 
the top of the band, in which case 
the divergence transforms into a strong enhancement.
We therefore expect that our numerical results will 
overestimate the strength of 
the low-frequency resonance. 

Finally, we shortly discuss some technical aspects of our numerical
calculations. It follows from Eq.(\ref{full_R}) that the expression for the Raman 
intensity contains four-dimensional integrals with strong singularities.
In order to make a numerical evaluation possible, one has to include a fermionic 
damping, which cuts the singularities.
 However, if the damping is too 
large, subleading terms become stronger 
than the triple resonance effect. We found, for example, that a fermionic 
damping $\Gamma=0.4J$, which was used in Ref.\cite{SchKam} almost destroys the resonance at 
$\omega_{res}^{(2)}$.
We therefore only consider relatively small fermionic dampings with 
$0.05J \leq \Gamma \leq 0.10J$. Furthermore, in order to ensure sufficient 
accuracy of the results, we evaluated the necessary integrals on lattices up to 
$1000 \times 1000$ sites. We verified in each case 
that the convergence of the results was satisfactory.\\   

\subsection{Raman line shape in $B_{1g}$ geometry}
\label{shape}
\begin{figure} [t]
\begin{center}
\leavevmode
\epsffile{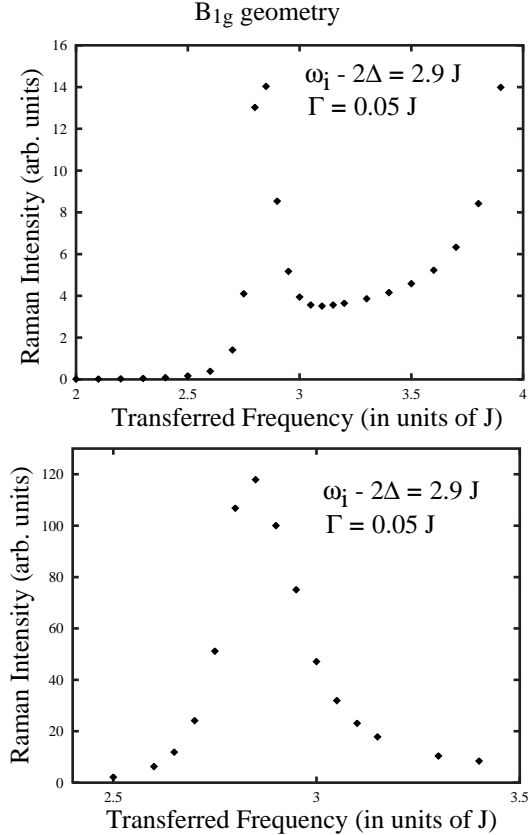}
\end{center}
\caption{The Raman intensity as a function of transferred frequency $\Delta \omega$ for
$\omega_i \approx \omega^{(1)}_{res}$  {\it a)} without and {\it b)} with a final state interaction.}
\label{b1g_1}
\end{figure}

We first present our numerical results for the Raman lineshape 
as a function of $\Delta \omega$ for fixed $\omega_i$. Our main result
is that the Raman lineshape evolves with increasing 
$\omega_i$ from a slightly asymmetric form 
at $\omega_i \approx \omega_{res}^{(1)}$ to a strongly asymmetric form at
$\omega_{res}^{(1)} < \omega_i < \omega_{res}^{(2)}$, and then back  to an
 almost symmetric form at 
$\omega_i \approx \omega_{res}^{(2)}$. In order to show this we 
present the results 
for three incident frequencies: $\omega_i \approx 2\Delta +2.9J$, 
$\omega_i \approx 2\Delta + 6J$, and
$\omega_i \approx 2\Delta + 7.9J$. In the first and third case  the 
triple resonance and the two-magnon peak positions coincide, whereas in the second case 
they are well separated.\\
{\it a)} $\omega_i \approx \omega^{(1)}_{res}$ \\
The Raman intensity as a function of transferred frequency without and with a final 
state interaction is presented in Fig.~\ref{b1g_1}a,b, respectively. 
The main difference 
between the two figures is the presence of an unphysical singularity 
in Fig.~\ref{b1g_1}a at $\Delta 
\omega_{max}=4J$ which is due to a divergent density of states at the boundary 
of the magnetic Brillouin zone. As in the LF-theory, 
the inclusion of a 
magnon-magnon interaction eliminates this singularity as is seen in 
Fig.~\ref{b1g_1}b. A more relevant point is that
 both figures contain a strong peak at $\Delta 
\omega=2.8J$.
 While the peak in Fig.~\ref{b1g_1}a is solely due to the 
divergence in the Raman vertex at $\omega^{(1)}_{res}$, 
the peak in Fig.~\ref{b1g_1}b 
is a combined effect of the 
resonance in the Raman vertex
and multiple magnon-magnon scattering.
We see that the 
peak in Fig.~\ref{b1g_1}a is strongly enhanced by the final state interaction.

\begin{figure} [t]  
\begin{center}
\leavevmode
\epsffile{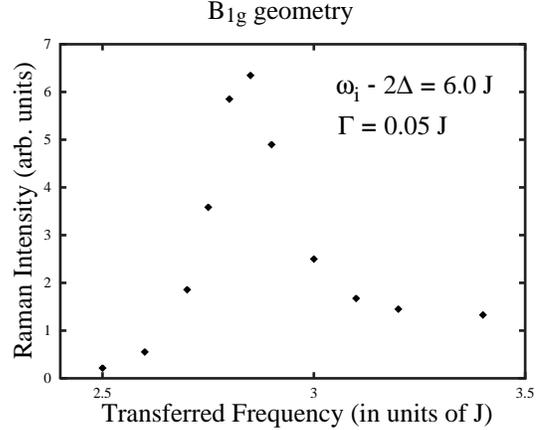}
\end{center}
\caption{The Raman intensity as a function of transferred frequency $\Delta \omega$ for 
 $\omega_i-2\Delta=6.0J$ in the interacting case. }
\label{b1g_3}
\end{figure}

Furthermore, we see
 that the Raman lineshape in Fig.~\ref{b1g_1}b is 
slightly asymmetric
with a larger intensity at higher transferred frequencies. 
This asymmetry is most likely to be a  
property of the Raman vertex since the
final state interaction yields a symmetric peak. One can indeed see 
this asymmetry already in Fig.~\ref{b1g_1}a. 
The two-magnon lineshape obtained numerically
is consistent with our analytical results in Sec.~\ref{om_res1}. There
we attributed the asymmetry of the two-magnon profile to the 
branch cut in the Raman vertex which for  $\omega_i = 2\Delta +2.9J$
exists only for $\Delta \omega > 2.8J$.\\
{\it b)} $\omega^{(1)}_{res}<\omega_i<\omega^{(2)}_{res}$ \\ 
The form of the Raman profile changes quite 
strongly as one moves from $\omega^{(1)}_{res}$ 
to intermediate incident frequencies. 
In Fig.~\ref{b1g_3} we present, as an example, the Raman intensity
including a final state interaction for $\omega_i=2\Delta+6.0J$. 
A comparison with 
Figs.~\ref{b1g_1} shows that the anisotropy of the intensity is now 
much stronger. This result is quite expected since in this range of incident frequencies, 
the Raman vertex resonates at transferred frequencies above the two-magnon peak.
In particular, for $\omega_i=2\Delta+6J$,
 the triple resonance occurs near the maximum
transferred frequency $\Delta \omega =4J$ (see Fig.~\ref{trip_res}). 

\begin{figure} [t] 
\begin{center}
\leavevmode
\epsffile{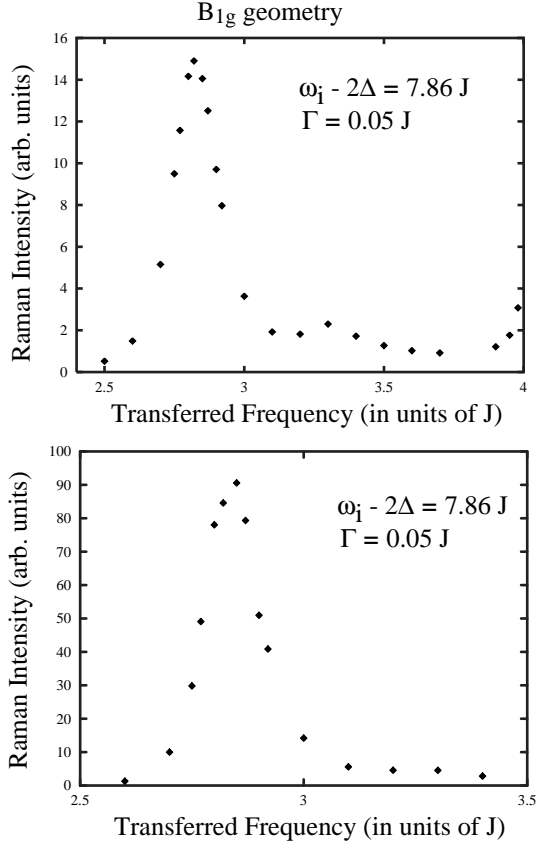}
\end{center}
\caption{The Raman intensity as a function of transferred frequency $\Delta \omega$ for 
 $\omega_i=\omega_{res}^{(2)}$, {\it a)} without and {\it b)} with a final state interaction.}
\label{b1g_2}
\end{figure}

The two-magnon profile for intermediate frequencies within the triple resonance
theory was earlier obtained by Sch\"{o}nfeld {\it et al.}. Our results are
in full agreement with theirs.\\
{\it c)} $\omega_i \approx \omega^{(2)}_{res}$ \\
The results for the intensity with and without final state interaction are
presented in Fig.~\ref{b1g_2}b,a, respectively. 
The intensity without a final state interaction
again exhibits an unphysical divergence at the maximum transferred frequency 
$\Delta \omega =4J$ which disappears when one
includes a magnon-magnon scattering. Near $\omega^{(2)}_{res}$, however, 
this divergence is confined to a very narrow region near $4J$.\\
Furthermore, we obtain that in both figures the
peak at around $\Delta \omega = 2.8J$ 
is almost symmetric.  This is
also consistent with our analytical results in Sec.~\ref{om_res2}.\\
In addition to the peak at $\Delta \omega = 2.8J$, 
both intensities also possess a slight maximum around $3.3 J$ which
probably originates from subleading, branch cut terms in the intensity.\\   

The evolution of the Raman profile with increasing $\omega_i$ 
from slightly asymmetric form around 
$\omega_{res}^{(1)}$
 to a pronounced shoulder-like behavior for intermediate 
frequencies, to a symmetric form close to $\omega_{res}^{(2)}$ is fully
consistent with
the experimental results on $Sr_2CuO_2Cl_2$ and $YBa_2Cu_3O_{6.1}$. 
We consider this agreement 
with the data as yet another evidence that the triple resonance diagram dominates the 
scattering process in the resonance regime.

\subsection{Two-magnon peak height}
\label{tmph}                          
\begin{figure}  [t]
\begin{center}
\leavevmode
\epsffile{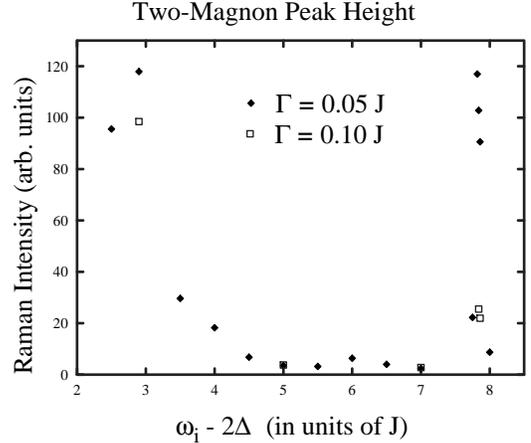}
\end{center}
\caption{The TMPH as a function of incident frequency $\omega_i$.}
\label{TMPH}
\end{figure}

We now discuss the TMPH as a function of $\omega_i$.
For the calculation of the  TMPH, we fix the transferred frequency at a value which corresponds 
to the maximum of the two magnon profile (which, depending on $\omega_i$, occurs 
between $\Delta \omega=2.8J$ and $\Delta \omega=2.9J$) and plot the intensity
of the maximum as a function of $\omega_i$. We present the results in 
Fig.~\ref{TMPH} for two different values of the fermionic damping $\Gamma$. In both cases 
we clearly observe two maxima at $\omega^{(1)}_{res} \approx 2.9J$ and 
$\omega^{(2)}_{res} \approx 7.9J$. 
The positions of these maxima 
are in good agreement 
with the analytical predictions and the experimental data.
The form of the TMPH near $\omega^{(1)}_{res}$ is clearly asymmetric:
the intensity drops faster above the peak than below. This form 
agrees with our analytical results.
For intermediate incident frequencies 
($4.0J<(\omega_i-2\Delta)<7.5J$) the TMPH 
remains basically constant and, in addition, 
is practically $\Gamma$-independent. 
This behavior, we believe, results from the fact  
that in this frequency range the triple resonance occurs at $\Delta \omega = 4J$ 
which is too far away from the two-magnon peak to influence its height.\\
Upon increasing $\Gamma$,
we find that the TMPH around $\omega_{res}^{(2)}$ drops  much more rapidly than 
around $\omega_{res}^{(1)}$. This is fully
consistent with our analytical result that the divergence 
in the Raman vertex, which is only cut by the fermionic damping, 
is present only near $\omega_{res}^{(2)}$
while the maximum near $\omega_{res}^{(1)}$ is just an enhancement which 
does not crucially depend on the damping.
 
\begin{figure}  [t]
\begin{center}
\leavevmode
\epsffile{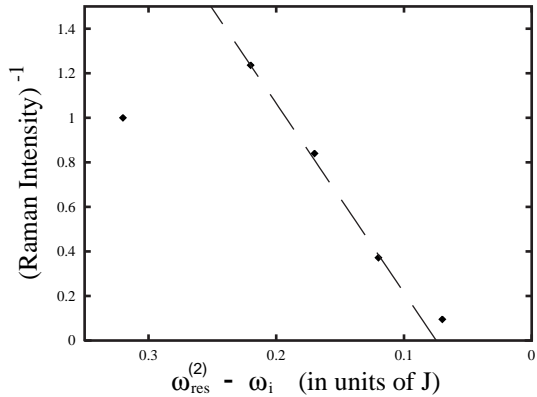}
\end{center}
\caption{The inverse Raman intensity for $\Delta \omega = 2.9J$ as a function of 
$\omega^{(2)}_{res} - \omega_i$.}
\label{inv}
\end{figure}

We also obtained two results which are not fully consistent with the
experimental data. The first one is the ratio of intensities at the two maxima.
We found that while the divergence in $M_R$ exists only near $\omega_{res}^{(2)}$, 
the non-divergent terms are much stronger around $\omega_{res}^{(1)}$. As a result, 
the ratio of intensities $I(\omega_{res}^{(1)})/I(\omega_{res}^{(2)})$
for $\Gamma=0.05J$ is $ \approx 1$, 
while experimentally, this ratio 
is clearly smaller than one, though the actual number differs between 
$Sr_2CuO_2Cl_2$ and $YBa_2Cu_3O_{6.1}$. 
The second discrepancy concerns  the behavior of the TMPH 
in the vicinity of
$\omega^{(2)}_{res}$. Analytically, we found that the TMPH 
should follow an inverse linear behavior at some distance from
$\omega^{(2)}_{res}$ and an inverse 
cubic behavior in the immediate
vicinity of $\omega^{(2)}_{res}$.
To verify this result, we calculated
 the TMPH for several frequencies in the vicinity of 
$\omega^{(2)}_{res}$ and present the results in Fig.~\ref{inv}
(the dashed line in this figure is 
a guide to the eye). Within our numerical accuracy,
 we indeed found an inverse linear dependence which,
however, only exists for a small region near $\omega^{(2)}_{res}$, 
namely for $ 0.1J < \omega^{(2)}_{res} - \omega_i < 0.25J$. 
Experimentally, this region extends 
over a much wider frequency range of about $1 eV$. 
Very close to $\omega^{(2)}_{res}$, the divergence is cut by the 
fermionic damping, and it is  impossible to verify
the predicted inverse cubic behavior.\\
We already mentioned
 that one of the reasons for the incorrect ratio of intensities
$I(\omega_{res}^{(1)})/I(\omega_{res}^{(2)})$ lies 
in the oversimplified mean-field form 
of the fermionic dispersion, and, in particular, in the degeneracy along the boundary 
of the magnetic Brillouin zone.  
One would thus expect a better agreement with experiments
if this degeneracy is lifted e.g., by the introduction of a finite next-nearest hopping
$t^\prime$. We address this issue in the next subsection.

\subsection{Raman intensity for a nonzero $t^{\prime}$}
\label{tprime}
\begin{figure}  [t]
\begin{center}
\leavevmode
\epsffile{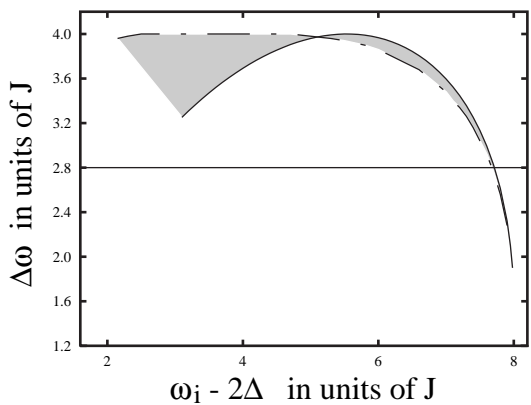}
\end{center}
\caption{The shaded area describes the region of the $(\omega_i,\Delta \omega)$ plane in which a 
double resonance occurs for $\omega_i - 2 E_k = 0$ and 
$\omega_i -{\Delta \omega \over 2}+ E_{k-q}^v - E_k^c = 0$. The solid line corresponds 
o $q_x=q_y$ and the dashed line to $q_x=0$.}
\label{dr}
\end{figure}

\begin{figure} [t] 
\begin{center}
\leavevmode
\epsffile{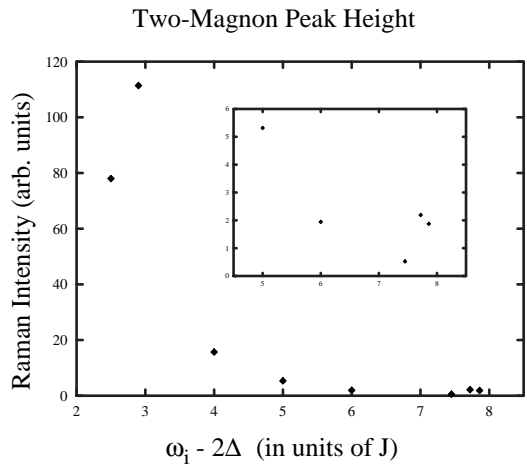}
\end{center}
\caption{The TMPH as a function of 
the incident frequency $\omega_i$ for $t^{\prime}/J=-0.3$ and $\Gamma=0.05J$.
The inset 
shows that despite the strong reduction one can still observe a maximum around 
$\omega_{res}^{(2)}$.}
\label{tmph_tp}
\end{figure}

In Sec.~\ref{om_res2a} we found 
that a nonzero $t^\prime$ splits the triple resonance around 
$\omega^{(2)}_{res}$ into three double resonances
 one of which occur in the same region of the 
$( \omega_i, \Delta \omega)$ plane as the triple resonance in the absence 
of $t^{\prime}$ while the other two occurs in different regions of the
$( \omega_i, \Delta \omega)$ plane. 
In Fig.~\ref{dr} we plot the 
region of the $( \omega_i, \Delta \omega)$ plane in which one of the remaining double 
resonances occurs. The form of the shaded area 
is similar to Fig.~\ref{trip_res}. 
We see that in the  vicinity of $\omega^{(2)}_{res}$
 the new resonant region  reduces to a
single line, just as the resonance for $t^{\prime} =0$.
Near $\omega_{res}^{(1)}$ the situation is more complex since the different double
resonances overlap.\\
In Fig.~\ref{tmph_tp} we present the result for
 the TMPH for $t^{\prime}/t=-0.16$.
A comparison with Fig.~\ref{TMPH} for the TMPH at $t^{\prime}=0$
shows that a finite  $t^{\prime}$ 
reduces the TMPH at both resonant frequencies $\omega^{(1,2)}_{res}$. 
This reduction is fully consistent with our analytical calculations
since now only two terms in the denominator of the Raman vertex vanish 
simultaneously while the third scales as $O(t^\prime/t)$. 
The reduction, however, 
is not uniform, and the TMPH around the high frequency resonance 
decreases much more 
rapidly than the one around the low-frequency resonance. 
Most probably, the increase of the ratio is caused by two effects. First, 
the regions of double resonance overlap around $\omega^{(1)}_{res}$,
but not around $\omega^{(2)}_{res}$. Second, 
a nonzero $t^\prime$ also affects the interaction vertex $V_{lf}$ 
between light and fermions and reduces it much 
more strongly around $\omega_{res}^{(2)}$ than 
around $\omega_{res}^{(1)}$. To see this, we recall 
that in the mean-field approximation 
we have $V_{lf} = (\partial \epsilon_k / \partial \vec{k}) \  \hat{e}_{i,f}$ with 
$\epsilon_k =-2t(\cos k_x + \cos k_y) + 4|t^{\prime}| \cos k_x \cos k_y$.
Near $\omega^{(2)}_{res}$,  the dominant contribution to the Raman vertex
comes from fermions near the bottom of the band (${\bf k} \approx 0$) in which case the 
vertex between light and fermions is reduced
by a factor of $(1-2|t^{\prime}|/t)$.  In contrast, the resonance near
$\omega^{(1)}_{res}$ is dominated by fermions near the top of the valence band
($k = (\pm \pi/2,\pm \pi/2))$ in which case the effect 
of $t^\prime$ is negligible.

 We see therefore 
that the inclusion of a finite $t^\prime$ actually 
worsens the agreement with the 
experiments since the ratio of intensities increases.
In the next subsection we will consider whether 
vertex corrections  can possibly 
reverse the effects of $t^{\prime}$ and restore the correct quantitative behavior of the 
TMPH. 

\subsection{Vertex Corrections}
\label{vertex}

\begin{figure} [t]
\begin{center}
\leavevmode
\epsffile{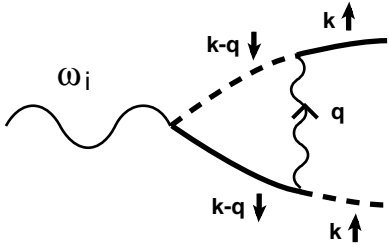}
\end{center}
\caption{The lowest order vertex corrections to the light-quasiparticle interaction. The 
solid and dashed lines represent the conduction and valence band quasiparticles, respectively.
The internal wavy line describes the exchange of a magnon.}
\label{vert_cor}
\end{figure}

There are several vertices in the diagram for the Raman matrix element, 
each of which is renormalized by vertex corrections which are generally not small
at large $U$. The calculation of all vertex
corrections is beyond our computational abilities and in this section we will therefore 
focus on the corrections to the vertex 
between light and fermionic 
quasiparticles, $V_{lf}$. 
Some evidence that the vertex between light and fermions near
$\omega^{(2)}_{res}$ is larger than in the
mean-field theory comes from 
the measurements of  the optical conductivity in 
$Gd_2CuO_4$, $Pr_2CuO_4$ and 
$YBa_2Cu_3O_6$ \cite{LiuKle}. 
These experiments have demonstrated that  the measured conductivity is 
larger than the one calculated with the mean-field form for $V_{lf}$ even 
though it basically follows the same frequency dependence. We will study the 
vertex corrections to $V_{lf}$ semi-phenomenologically and 
our goal will be to illustrate how  they can, in principle, 
reverse the effects of $t^{\prime}$.\\
The lowest order  correction to $V_{lf}$ in a formal perturbative 
expansion in $1/S$ is presented in Fig.~\ref{vert_cor}.
A simple analysis shows that the relative
vertex correction scales as $U/JS$, i.e., it is small only in the limit
of a very large spin. For realistic $S$, however, we have $U/JS \gg 1$,
and the corrections to $V_{lf}$ are large. This clearly implies that 
one should sum up an infinite series of corrections
to obtain the proper renormalization of the vertex between light and fermions.
We will not do this but rather model the effect of the vertex corrections phenomenologically by 
introducing an effective vertex in the form 
\begin{equation}
V^{eff}_{lf}(k) = \Big( \displaystyle{\partial \epsilon_k
\over \partial \vec{k}} \hat{e}_{i,f} \Big) \Big(1+\alpha \; \nu_k^2 \Big)   
\label{full}
\end{equation}              
with $\alpha$ as a parameter.
The effective vertex in Eq.(\ref{full}) still 
possesses the same symmetry as the bare vertex $V_{lf}(k)$ 
and therefore still 
vanishes at the bottom of the band. However, the 
slope of $V^{eff}_{lf}(k)$ around ${\bf k}=0$ 
can now be quite different from the mean-field result.

\begin{figure} [t]
\begin{center}
\leavevmode
\epsffile{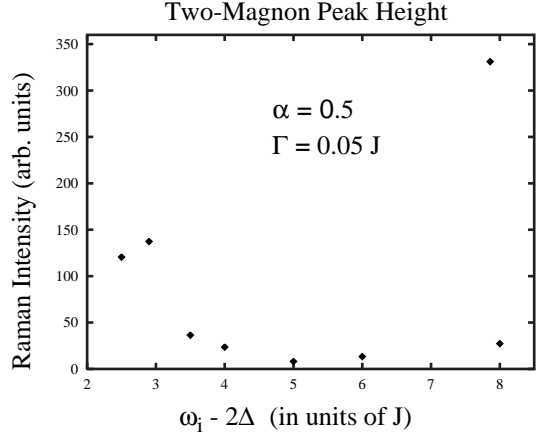}
\end{center}
\caption{The TMPH for $\Gamma=0.05J$ and  $\alpha=0.5$.}
\label{vc_a}
\end{figure}

We computed the TMPH with the effective vertex using various values of $\alpha$.
The result for $\alpha = 0.5$ is presented in Fig.~\ref{vc_a}.
A comparison with Fig.~\ref{TMPH} shows that the effect of the vertex correction is rather 
strong; the ratio $I(\omega^{(1)}_{res})/I(\omega^{(2)}_{res})$ 
is decreased 
by a factor of about $2.5$. 
In addition, we also observe a relative increase 
of the TMPH for intermediate $4.5J \leq (\omega_i-2\Delta)  \leq 5.5J$. 
This  last effect leads to an extension 
of the region in which the Raman intensity possesses an inverse linear 
behavior.

The decrease of the ratio of the intensities and the extension
of the frequency range of the inverse linear behavior
are both in  agreement with the experimental results~\cite{BluAbb}. 
We therefore see that by adjusting the 
vertex corrections to $V_{lf}(k)$ without violating the symmetry requirements 
of the model, one can, 
in principle, not only obtain good qualitative, but also quantitative 
agreement with the experimental data. The question is, however,  whether, e.g., 
$\alpha =0.5$ which we used in Fig.~\ref{vc_a} can be obtained in a microscopic
calculation. These studies are clearly called for.

\section{Discussion}
\label{discussion} 
We first summarize our results. The intent of this 
paper was to study the full Raman 
intensity in the resonant regime by simultaneously
 considering the effects of the 
triple resonance in the Raman vertex and  
the final state magnon-magnon interaction. 
We derived an explicit expression for the full Raman intensity
in the resonant regime as a function of both, 
transferred frequency $\Delta \omega$ and incoming
frequency $\omega_i$.  
We obtained analytically and numerically the
two-magnon Raman profile as a function of the 
transferred photon frequency $\Delta\omega$
and the dependence of the two-magnon peak height on the 
incident photon frequency
$\omega_i$. We found that the resonant behavior
 of the Raman vertex survives the inclusion of a 
magnon-magnon interaction and obtained two maxima in the peak height at 
$\omega_{res}^{(1)} \approx 2\Delta + 2.9J$ and at
$\omega_{res}^{(2)} \approx 2\Delta + 7.9J$. The position of 
the two maxima are the same as
in the semi-phenomenological approach by CF which considered the triple resonance enhancement
and final state interaction independent of each other.
We first studied in detail the two-magnon profile at various incident
frequencies. We found that the two-magnon peak is slightly asymmetric near
$\omega_{res}^{(1)}$ with larger intensity at higher frequencies.
As the incident frequency increases,
the asymmetry becomes stronger, and the two-magnon profile acquires a shoulder-like
feature above the peak. 
This is consistent with earlier results~\cite{SchKam}. 
For frequencies around $\omega_{res}^{(2)}$,
however, we found that the anisotropy disappears, and the
Raman profile acquires almost the same form as in the nonresonance, LF regime.

We then proceeded to a more detailed study of the two-magnon peak height.
We verified that the inverse linear behavior of the Raman intensity
near $\omega^{(2)}_{res}$  survives the effect of the final state
interaction. Furthermore, we considered the behavior of the Raman vertex near 
$\omega_{res}^{(1)}$. We found 
in our analytical considerations for which we assumed an isotropic dispersion near the top of
the band that the divergence is almost completely 
suppressed. However, the Raman vertex contains a branch cut which gives rise to an 
enhancement of the
intensity in some range of frequencies $\omega_i \leq \omega_{res}^{(1)}$
which terminates only slightly below $\omega_{res}^{(1)}$. 
In our numerical calculations, for which 
we considered a mean-field form of the dispersion, we obtained a
weak singularity at $\omega_{res}^{(1)}$ but also 
a strong enhancement of the Raman intensity for
$\omega_i$ slightly smaller than $\omega_{res}^{(1)}$. 
This last enhancement is virtually independent of the damping.
 
We found that the ratio of the Raman intensities
$I(\omega^{(1)}_{res})/I(\omega^{(2)}_{res})$ is already rather large 
for small damping, contrary to the assertion by CF. A much smaller ratio is
needed for a quantitative agreement with the experimental data.
 We attribute the 
large ratio to an unexpectedly strong enhancement of the two-magnon peak near 
$\omega^{(1)}_{res}$  due to a branch-cut anomaly in the Raman vertex.
  
We further studied how the  the triple resonance is modified by 
 a next-nearest hopping term $t^{\prime}$. 
 Around $\omega_{res}^{(2)}$, 
 we found that the triple resonance is split into 
three double resonances, but the linear divergence of
 the Raman intensity near $\omega_{res}^{(2)}$  is not changed. 
This splitting, however, reduces the 
intensity around $\omega_{res}^{(2)}$
relative to the intensity around $\omega_{res}^{(1)}$
 where the effect of a finite $t^\prime$
is rather weak. As a result, the ratio of the intensities 
$I(\omega^{(1)}_{res})/I(\omega^{(2)}_{res})$ increases.

Finally, we have demonstrated that the ratio of the 
intensities at the two resonance values of
$\omega_i$ is sensitive to the actual form of the vertex between light
and fermions. We have shown that the corrections to the 
mean-field vertex are large
and modeled their effect by introducing an extra 
factor $(1 + \alpha \nu_k^2)$ 
into the vertex.
We considered $\alpha$ as an adjustable parameter and showed that 
the ratio of the intensities can be substantially reduced already for moderate values of $\alpha$.

We now discuss our results in the context of the key experimental features that we listed in 
the introduction as being in disagreement with the LF theory:\\
1) {\it Changing lineshape with} $\omega_i$\\
Our results for the evolution of the Raman profile with $\omega_i$ is in complete agreement 
with the experimental results by Blumberg {\it et al.} \cite{BluAbb} on $YBa_2Cu_3O_{6.1}$. 
For $Sr_2CuO_2Cl_2$, the 
highest experimentally accessible frequency is smaller
 than $\omega_{res}^{(2)}$ and we therefore do not know whether 
the Raman profile eventually becomes symmetric near $\omega_{res}^{(2)}$. 
For intermediate $\omega_i$, however, our results agree with the experimental data.

An issue which  we have not addressed in our approach is the actual rather
than relative width of the two-magnon peak.  Experimentally, it is much broader than in 
our model. 
Previous studies by Weber and Ford \cite{WebFor}, however, 
have shown that the broadening may be due to a magnon damping. They
demonstrated that a small  damping due to, e.g., an interaction with phonons 
already gives rise to a considerable broadening of the two-magnon peak.
This result has also been obtained in numerical studies \cite{Haas}.

Another reason for the broadening of the Raman lineshape is the fermionic damping 
$\Gamma$. Incidentally, this damping may also account for the experimentally 
observed additional broadening of the TMPH around $\omega_{res}^{(2)}$ since in this frequency 
range, the dominant contribution to $M_R$ comes from fermions with ${\bf k} \approx 0$ which 
exhibit the largest damping \cite{BluAbb}.
 
2) {\it The TMPH as a function of} $\omega_i$\\
The two key experimental results for the TMPH, we remind, are the presence of two maxima 
in the TMPH, of which the higher frequency maximum
is stronger in all compounds, and an inverse linear behavior of the Raman 
intensity near the upper resonance frequency $\omega^{(2)}_{res}$.
In our 
analytical and numerical calculations we found the two maxima in the TMPH
whose positions fully agree with the experimental data.  In addition we found that the 
low-frequency maximum in the TMPH is anisotropic with a higher intensity at lower frequencies
which is also consistent with the experimental results.

Our numerical data, however, differ quantitatively from the experimental results in that 
the ratio $I(\omega_{res}^{(1)})/I(\omega_{res}^{(2)})$ is too large. 
On the basis of our analytical results we would expect the opposite behavior since 
we found that the actual resonance in the Raman intensity 
(i.e. a divergence in the absence of a fermionic damping)
exists only near $\omega^{(2)}_{res}$ while the peak near $\omega^{(1)}_{res}$
is just the enhancement due to nonsingular terms in the Raman vertex.  It turns out however
that these nonsingular terms are anomalously large.
Naively, one would expect that the inclusion of a next-nearest neighbor 
hopping $t^{\prime}$ 
would lead to an improved agreement of the ratio with the experimental data.
In contrast, we found that a finite $t^{\prime}$ suppresses the 
high-frequency resonance even further.
On the other hand,  we have demonstrated that the inclusion of the 
corrections to the interaction between light and  fermions 
may substantially increase the vertex near $\omega^{(2)}_{res}$ compared to the
vertex near $\omega^{(1)}_{res}$. This eventually yields a
much better ratio of $I(\omega_{res}^{(1)})/I(\omega_{res}^{(2)})$
which can be made fully consistent with the experimental data by adjusting the magnitude of
the vertex correction.

Our analytical and numerical computations also reproduced the inverse linear behavior of the 
Raman intensity, which was observed in $YBa_2Cu_3O_{6.1}$ \cite{BluAbb} and 
$PrBa_2Cu_3O_7$ \cite{RubDie} and, to a certain extent, also in $Sr_2CuO_2Cl_2$. However,
we also found that the range of $\omega_i$ in which this 
behavior was observed experimentally 
is much larger than in our analysis. The inclusion of the vertex corrections improves the
agreement with the data but does not make it perfect.
This issue requires further study.

In conclusion we have provided a detailed study of Raman scattering in the resonant regime. 
We confirmed that the key experimental features of magnetic Raman scattering can be explained 
qualitatively, and to some extent quantitatively within the 
triple resonance theory. We believe that the remaining quantitative discrepancies are due to  
insufficient knowledge of the quasi-particle energy dispersion, lifetime effects and the form 
of the vertex function between light and fermions.

\begin{figure} [t] 
\begin{center}
\leavevmode
\epsffile{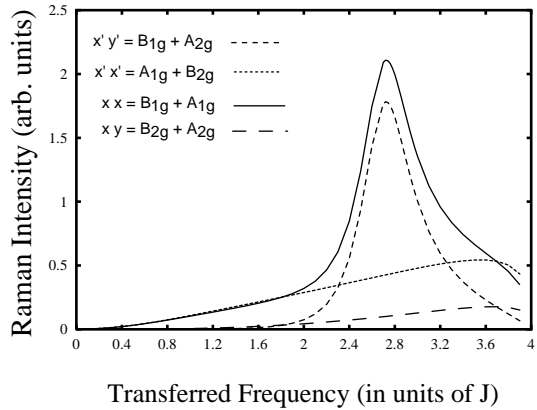}
\end{center}
\caption{The Raman intensity as a function of transferred frequency $\Delta \omega$ for
 constant denominator.}
\label{numerator}
\end{figure}

\begin{figure} [t] 
\begin{center}
\leavevmode
\epsfxsize=7.5cm
\epsffile{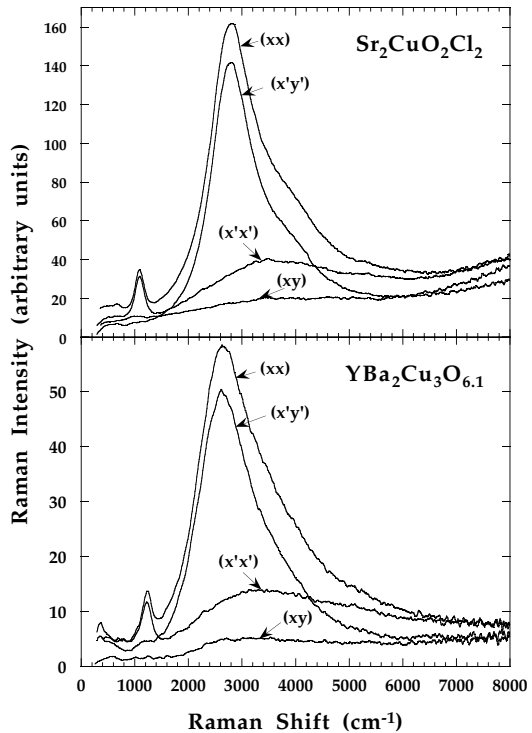}
\end{center}
\caption{
Two-magnon Raman scattering spectra from Sr$_2$CuO$_2$Cl$_2$ and 
YBa$_2$Cu$_3$O$_{6.1}$ at room 
temperature in different scattering geometries. 
 Courtesy of the authors of Ref.~\protect\cite{Blu2}. The labels indicate:
$xx$ ($B_{1g} + A_{1g}$),
$x'y'$ ($B_{1g} + A_{2g}$),  
$x'x'$ ($A_{1g} + B_{2g}$) and 
$xy$ ($B_{2g} + A_{2g}$) 
The excitation energy is $2.73$~eV, which is about $0.4$~eV below $\omega_{res}^{(2)}$. }
\label{srcl}
\end{figure}

A final remark. We discussed in Sec.~\ref{shape} that
for intermediate incident frequencies, $\omega^{(1)}_{res} < \omega < \omega^{(2)}_{res}$,
the triple resonance occurs relatively far from the 
two-magnon peak and, to first approximation, does not influence the two-magnon lineshape.
In other words, in calculating the two-magnon lineshape, one can, with reasonable accuracy,
set the denominator in the triple resonance diagram 
to a constant and  compute the Raman vertex in the same 
way as in the LF theory. It turns out that this procedure yields good agreement
with the data not only in $B_{1g}$ but also in other scattering geometries.
To demonstrate this, we present 
in Fig.~\ref{numerator} the results for the intensity in four different 
scattering geometries. These results have to be
compared with  the experimental data for
$Sr_2CuO_2Cl_2$ and $YBa_2Cu_3O_{6.1}$ \cite{Blu2}. 
which we reproduce in Fig.~\ref{srcl}.  Since $J \sim 1000 cm^{-1}$, the
comparison
with Fig.~\ref{numerator} is valid only for $\Delta \omega \leq 4000 cm^{-1}$.
A finite scattering intensity at larger $\Delta \omega$ is probably due to
multi-magnon scattering.  
We consider the agreement between the two figures as 
rather good and view it as an
another piece of evidence in favor of  the triple resonance theory.\\[0.3cm]    

\newpage
\centerline{{\bf Acknowledgments}}
It is our pleasure to thank J. Betouras, G. Blumberg, D. Frenkel, R. Joynt
 and M.V. Klein for helpful
conversations. The research was supported by NSF-DMR 9629839.
A.C. is an A.P. Sloan fellow.

\appendix
\section{The Raman intensity with magnon-magnon interaction}
\label{full_raman} 
In this appendix we present the formulae for the numerical computation of the full Raman 
intensity. Our starting point is the Golden Rule formula,  Eq.(\ref{golden})
\begin{equation}
R(\omega_i,\Delta \omega) \propto   \int {d^2q \over 4 \pi^2} 
|M^{tot}_R(\omega_i,\Delta \omega,q)|^2 \delta(\Delta \omega-2 \omega_q)
\label{full_R}
\end{equation}
where $ M^{tot}_R$ is diagrammatically presented
 in Fig.\ref{diaint}b. The Golden Rule formula  for the 
intensity corresponds to the diagram in Fig.~\ref{diaint}a
 in which  the intermediate magnons are on the mass shell.\\
The analytical expression for $M_R^{tot}$ has the form
\begin{equation}
M^{tot}_R=M_R+ { \bar{M}_R \ \tilde{\nu}_q \over 1+I/4S }
\end{equation}
where $I$ is given in Eq.(\ref{eff_ver}), and 
$M_R$ and $\bar{M}_R$ are given by 
Eqs.(\ref{tripa}) and (\ref{trip}) for $t^{\prime} =0$ and by
Eqs.(\ref{tripa_tp}) and (\ref{trip_tp}) for $t^{\prime} \neq 0$, respectively.\\
Note that we use the full form of $M_R$ and do not project it on $\tilde{\nu}_q$ as was done in 
Ref.\cite{SchKam}. Our numerical computations show that especially for small $\Gamma$, 
the $B_{1g}$ component of $M_R$ has a more complex dependence on the magnon momentum than
just ${\tilde \nu}_q$.

\end{document}